\documentclass[aps,prb,twocolumn,superscriptaddress,showpacs,showkeys]{revtex4-2}  
\usepackage[pdftex]{graphicx}
\usepackage{color}
\usepackage{amsmath}
\usepackage{multirow}
\usepackage{placeins}

\usepackage{changes}

\usepackage{xr}
\begin{document}


\title{Nature of Interfacial Dzyaloshinskii-Moriya Interactions in Graphene/Co/Pt(111) Multilayer Heterostructures}




\newcommand{\QUIMI}[0]{{
Departamento de Pol\'{\i}meros y Materiales Avanzados: 
F\'{\i}sica, Qu\'{\i}mica y Tecnolog\'{\i}a, Facultad de Qu\'{\i}mica UPV/EHU,
Apartado 1072, 20080 Donostia-San Sebasti\'an, Spain}}

\newcommand{\DIPC}[0]{{
Donostia International Physics Center, 
Paseo Manuel de Lardiz\'abal 4, 20018 Donostia-San Sebasti\'an, Spain}}

\newcommand{\CFM}[0]{{
Centro de F\'{\i}sica de Materiales CFM/MPC (CSIC-UPV/EHU), 
Paseo Manuel de Lardiz\'abal 5, 20018 Donostia-San Sebasti\'an, Spain}}

\newcommand{\JUELICH}[0]{{
Peter Gr\"unberg Institut and Institute for Advanced Simulation, 
Forschungszentrum J\"ulich and JARA, 52425 J\"ulich, Germany}}

\newcommand{\ICMM}[0]{{
Instituto de Ciencia de Materiales de Madrid, CSIC,
Cantoblanco, 28049 Madrid, Spain}}

\author{M. Blanco-Rey}
\affiliation{\QUIMI}
\affiliation{\DIPC}
\author{G. Bihlmayer}
\affiliation{\JUELICH}
\author{A. Arnau}
\affiliation{\CFM}
\affiliation{\QUIMI}
\affiliation{\DIPC}
\author{J.I. Cerd\'a}
\affiliation{\ICMM}



\date{\today}

\begin{abstract}
DFT calculations within the generalized Bloch theorem approach show that 
interfacial Dzyaloshinskii-Moriya interactions (DMI) at both 
interfaces of Graphene/Co$_n$/Pt(111) multilayer 
heterostructures are decoupled for $n \geq 3$.
Unlike the property of magnetocrystalline anisotropy for this system, 
DMI is not affected by stacking defects in the Co layer.
The effect of Graphene (Gr) is to invert the chirality of the vaccum/Co interfacial DMI, 
overall reducing the DMI of the heterostructure, which is nevertheless 
dominated by the strong spin-orbit coupling (SOC) of Pt.
A spectral analysis in the reciprocal space shows that DMI at 
both the Gr/Co and Co/Pt interfaces have the same nature, 
namely SOC-split hybrid bands of $d$-orbital character. 
This proves that a DMI model based on a single band, 
such the Rashba DMI model, is insuficient to describe the behaviour 
of this family of Gr-capped $3d/5d$ metal heterostructures.
\end{abstract}


\maketitle

\section{Introduction}
\label{sec:intro}

When the magnetic exchange interactions are subject to sufficiently intense 
spin-orbit coupling (SOC) in an environment with broken inversion symmetry, 
an antisymmetric term appears that leads to canted and chiral orientation 
of spins, known as 
Dzyaloshinskii-Moriya interaction (DMI). 
Firstly observed for oxides \cite{bib:dzyaloshinskii57} and modelled 
as orbital-magnetic-moment-dependent terms added
to the Anderson hamiltonian \cite{bib:moriya60,bib:moriya60b}, 
it was later generalized to 
metallic alloys with diluted magnetic impurities (Fert-Levy model) 
as a SOC correction of the RKKY exchange \cite{bib:fert80,bib:fert81}.

Because of the broken inversion symmetry requirement, 
DMI is usually active at surfaces and interfaces, 
where it triggers complex chiral ordered spin structures at the nanoscale, 
such as cycloidal textures \cite{bib:bode07,bib:schmitt19} and
skyrmions \cite{bib:wiesendanger16,bib:fert17,bib:bogdanov20}, and introduces asymmetry in 
the displacement of domain walls \cite{bib:heide08}. 
The latter property has been exploited in synthetic magnets \cite{bib:luo19,bib:hrabec20}. 

Improving the efficiency of the chirality, which 
is lost if symmetric exchange interactions dominate, has been 
identified as a near-future challenge in the field of magnetic materials \cite{bib:roadmap20}.
Asymmetric multilayering is used to enhance DMI, since interfaces contribute 
additively \cite{bib:moreau16,bib:yang18b,bib:perini18}. 
For example, in Ir/Co/Pt- \cite{bib:moreau16} 
and  Pt/Co/Ta-based heterostructures \cite{bib:woo16} skyrmions of diameter $\sim 100$\,nm 
have been stabilized at room temperature.
Contact between a ferromagnet and a heavy non-magnetic phase is the obvious way to 
promote DMI, as hybridization with the spin-orbit split $5d$ bands 
facilitates the needed spin-flip electron excitations, hence
Co/Pt has become a paradigmatic system \cite{bib:yang15,bib:belabbes16,bib:sandratskii17}. 
Alternatively, DMI strength and handedness can also be manipulated by adsorption of 
light element atoms, , hydrogen \cite{bib:chen21,bib:yang20} and 
oxygen \cite{bib:belabbes16b,bib:chen20}, 
or capping with graphene (Gr) \cite{bib:ajejas18,bib:yang20} and 
hexagonal boron nitride \cite{bib:hallal21}. 
This DMI variation upon adsorption stems from the charge density redistribution 
at the surface and it is correlated with the electric dipole
at the surfaces \cite{bib:belabbes16b}, where the correlation 
is endorsed by an analytical expression \cite{bib:jia20}.
The manifestation of DMI through other more accessible properties of the system, 
such as electrostatic ones, has motivated the search of 
DMI descriptors that allow its predictability in systems of potential interest. 
For example, at a $3d/5d$ interface, it correlates with the spin moment $m_S$ 
of the $3d$ atoms but not with the $m_S$ induced at the interfacial $5d$ 
atoms \cite{bib:yang15,bib:belabbes16,bib:sandratskii17,bib:kim18b,bib:jia20}, 
neither with the spin dipole nor with the orbital magnetic moments \cite{bib:jia20}.
Direct modification of DMI merely based in electric field manipulation,  by a STM tip, is questionable, 
as it may be mistaken with a variation of the magnetic exchange coupling strength \cite{bib:hsu17}.

In an attempt to classify DMI into types, 
it has been argued that Co/Pt and Gr/Co interfacial DMI are of different nature, namely
Fert-Levy-like and Rashba-like, respectively \cite{bib:yang18}.
In the former interface, the DMI energy contribution localizes at Pt, which has 
of strong atomic SOC strength $\xi$. 
In the latter, Gr/Co, the aforementioned induced electrostatic potential change $\nabla V$ 
is modelled by a one-band Rashba hamiltonian in the presence of a Heisenberg exchange term 
\cite{bib:imamura04,bib:kim13,bib:kundu15}.
These models can be considered as limits of the strongly hybridized SOC-split $d$-bands behavior.

Motivated by recent experimental work on multilayer Gr/Co$_n$/Pt(111) 
heterostructures \cite{bib:ajejas18,bib:blanco21}
and by the existing density-functional theory (DFT) 
work on the individual interfaces of them, 
in this paper we show that that DMI 
does not adjust satisfactorily to either limit model. 
Our DFT calculations reveal that the 
Pt SOC dominates overall and that there are sizable contributions from Co SOC locally, 
although cancellations occur for Co at both interfaces. 
The DMI chirality induced by the graphene capping is opposite to
and of the same order of magnitude as that of the Co/Pt interface, in agreement 
with the observation of Ref.~\cite{bib:ajejas18}. 
We draw the important conclusion that this result cannot be 
simply attributed to specific states in the reciprocal space, 
i.e., a single-band Rashba model cannot account for it. 

During pseudomorphic growth by intercalation in Gr/Pt, Co attains a face tetragonal $fcc$ 
distorted ($fct$) stacking. Therefore, the central regions of the slab are locally 
centrosymmetric and DMI is solely an interfacial effect. 
In the present work we show that this regime is reached at Co$_3$ thickness and 
that the DMI contributions of the stacking defects cancel out.
Interestingly, this DMI behaviour contrasts with that of the 
magnetocrystalline anisotropy of $fct$ Gr/Co$_n$/Pt(111), 
where a complex behaviour with Co$_n$ thickness is found that 
depends on the competition between the contributions from the $fct$ bulk (in-plane)
and twin boundary defects (out-of-plane) \cite{bib:blanco21}. 

The paper is organized as follows:
in Section~\ref{sec:theory} we describe the model structures of Gr/Co$_n$/Pt 
used in this work and the details of the DFT calculations based on the generalized 
Bloch theorem with SOC. The results and discussion section is split into 
a collection of thickness-dependent and layer-resolved DMI energies 
(section \ref{sec:additivity}), as well as an analysis in the reciprocal 
space (section \ref{sec:reciprocal}), 
where the DMI contributions are energy- and wavevector-resolved. 
Finally, conclusions are drawn.

\section{Theoretical Methods} 
\label{sec:theory}

\begin{figure}[tb!]
\centerline{\includegraphics[width=1\columnwidth]{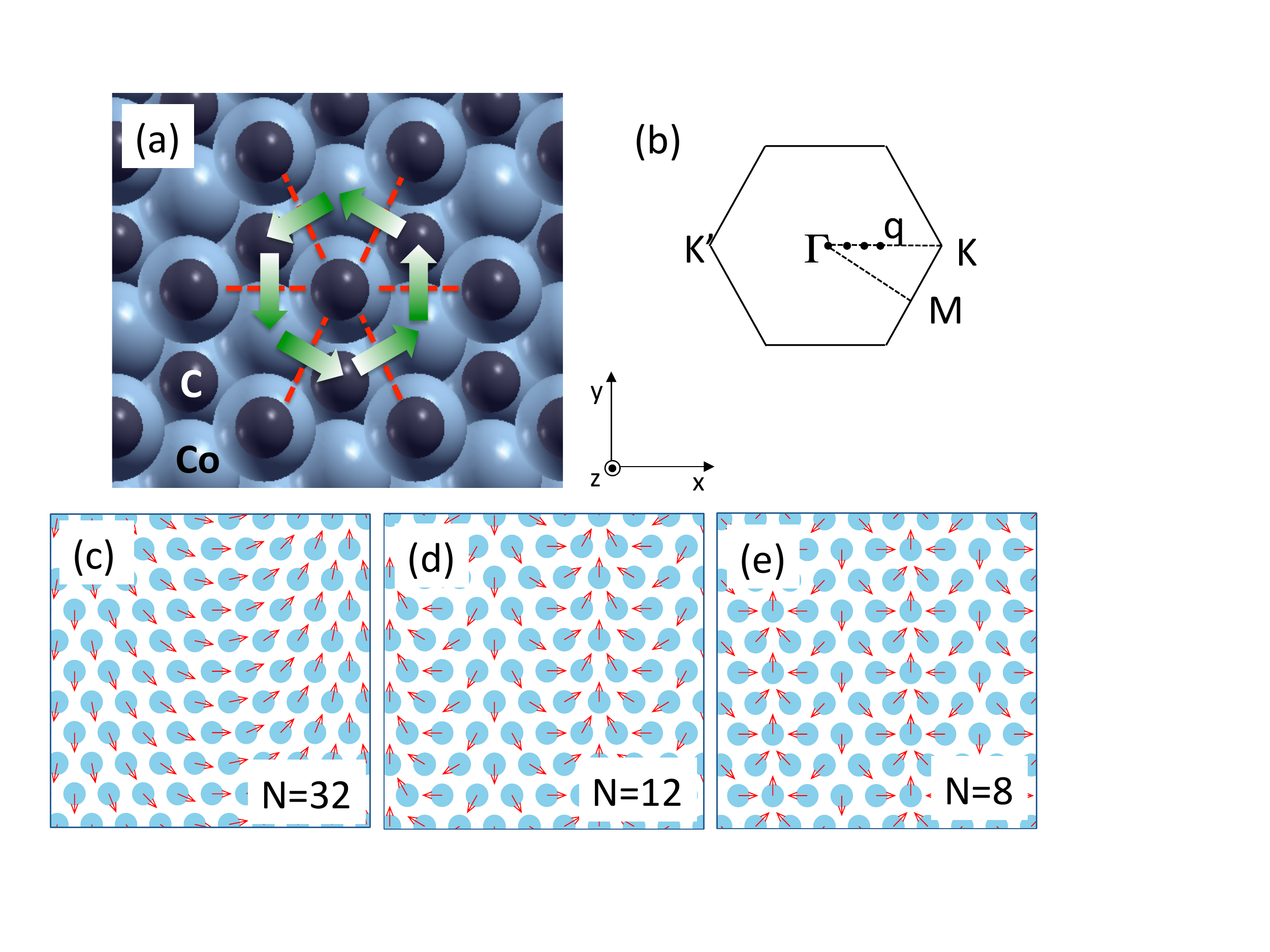}}
\caption{
(a) Top view of a graphene-covered $fct$ Co slab.
For the  topmost Co plane, the central atom $D$-vectors of the
DMI with its in-plane nearest neighbours (red dashed segments) are
indicated by thick arrows with 3-fold symmetry. The colour gradient from green to white
depicts the $D_z$ component sign.
(b) The spin spiral wavevectors used in this work belong to the $\Gamma K$
direction of the Brillouin zone, i.e.
$\mathbf{q} = \frac{2\pi}{a} ( \frac{1}{N}, \frac{1}{N}, 0)$.
(c-e) Examples of the spin orientations in a Co plane
for three $N$ values.
}
\label{fig:sketch}
\end{figure}

The pseudovectors $\mathbf{D}_{ij}$ characterize the DMI between two localized
spin moments $\mathbf{S}_i,\mathbf{S}_j$, which is expressed as the 
hamiltonian term $\sum_{\langle ij \rangle} \mathbf{D}_{ij} \cdot \mathbf{S}_i \times \mathbf{S}_j$.
Fig.~\ref{fig:sketch} shows the $D$-vectors between the Co atom at the origin
unit cell and its six closest Co atoms in the case of a monolayer. 
The 3-fold axes and mirror planes in the 
structure constrain these $D$-vectors to be determined by two free 
parameters $D_y$ and $D_z$ \cite{bib:crepieux98}, as indicated in the sketch.
Note that the relative positions of the nearest Pt atom with respect to 
each pair of Co neighbours alternates from right to left in the 3-fold symmetry 
and, thus, the $D$-vector out-of-plane component $D_z$ alternates in sign.
As this work is restricted to relatively small angles between spins,
we will use an effective model of the energy that 
maps all Co-Co interactions in the model slabs onto a two-dimensional 
hamiltonian that depends on effective $D_y$ and $D_z$ parameters (see Fig.~\ref{fig:sketch} 
and Supplemental Material Fig.~\ref{SM-fig:geom_Ds}).

DFT calculations in the full-potential linearized augmented
plane waves (FLAPW) formalism~\cite{bib:krakauer79,bib:wimmer81} 
have been carried out with the code FLEUR~\cite{bib:fleur}. 
PBE is the chosen exchange and correlation functional~\cite{bib:pbe96} 
for this work.
We have used Co layers of one to five atom thicknesses,
pseudomorphic on a Pt substrate (lateral periodicity
2.772\,{\AA}) of five atomic planes, with the  relaxed interplanar 
spacings found in Ref.~\cite{bib:blanco21}.
The effect of the substrate thickness on the DMI is shown in the 
Supplemental Material Fig.~\ref{SM-fig:pt_thickness}. 
For the basis set, a $48 \times 48 \times 1$ Monkhorst-Pack-point mesh~\cite{bib:monk76} 
is used and plane wave expansion cut-offs of 4, 11 and 9.5\,a.u. for the wavefunctions, 
density and potential, respectively. The local basis was constructed without local orbitals,
with $l_{max}=6,8,10$ for C, Co and Pt, respectively. 
The smearing energy for the Fermi level determination was 0.03\,eV.

Suitable non-collinear spin structures are needed
for the model to show DMI. They are modelled as long-wavelength 
spin spirals in the generalized Bloch theorem (GBT) 
approach \cite{bib:sandratskii91}, which imposes a 
longer periodicity of the magnetization density described 
by a wavevector $\mathbf{q}$. 
Although, in principle,
the charge density can be self-consistently converged 
for a given $\mathbf{q}$, since the calculations shown here involve 
long-wavelength spirals, we use the GBT non-self-consistently 
to calculate new energies and electron wavefunctions 
as a perturbation of a $\mathbf{q}=0$ ground state (see Supplemental Material Fig.~\ref{SM-fig:Eq}). 
Our calculations on the minimal model system, a Gr/Co$_1$/Pt$_1$ 
trilayer, show that this is a fair approximation in the low-$q$ regime
(see Supplemental Material Fig.~\ref{SM-fig:nscf_vs_scf}).
Finally, spin-orbit interactions are added also as 
a first-order perturbation. The implementation of this procedure 
in the FLAPW code is described in Ref.~\cite{bib:heide09}.

We use flat spin spirals with a spin rotation axis $\hat s_a$, 
which yields a different energy in the presence of SOC. 
The energy difference between axes pointing in opposite 
directions $\hat s_a^+ (\hat s_a^-)$ 
is the DMI energy for a given spiral with  $\mathbf{q}$ periodicity.
Our convention is that the spirals are anticlockwise (ACW) 
with respect to  $\hat s_a$, so that $\hat s_a^+ (\hat s_a^-)$ mean
ACW (CW) or left-handed (right-handed) spirals. Therefore, if the energy difference
$\Delta E_{DMI}^{\hat s_a}(\mathbf{q}) = E_{DMI}(\mathbf{q}; \hat s_a^+) - E_{DMI}(\mathbf{q}; \hat s_a^-)$
is positive (negative), a CW (ACW) spiral is favoured.  
The FLAPW calculations of this work are run 
for spirals with $\mathbf{q}$ vectors along the 
$\Gamma K$ direction of the two-dimensional first Brillouin zone
$\mathbf{q} = \frac{2\pi}{a} ( \frac{1}{N}, \frac{1}{N}, 0)$,
direction in the Brillouin zone. As an example, Fig.~\ref{fig:sketch}
shows the spins of a Co atomic plane in the cases of $N=32,12,8$ and $\hat s_a = Z$. 
We used $N$ values between 8 and 48 (the latter is at the resolution limit 
marked by our electron momentum $\mathbf{k}$ calculation grid, $48 \times 48 \times 1$).

\section{Results and discussion}
\label{sec:results}

\subsection{Additivity of interfacial DMI}
\label{sec:additivity}

\begin{figure}[tb!]
\centerline{\includegraphics[width=1\columnwidth]{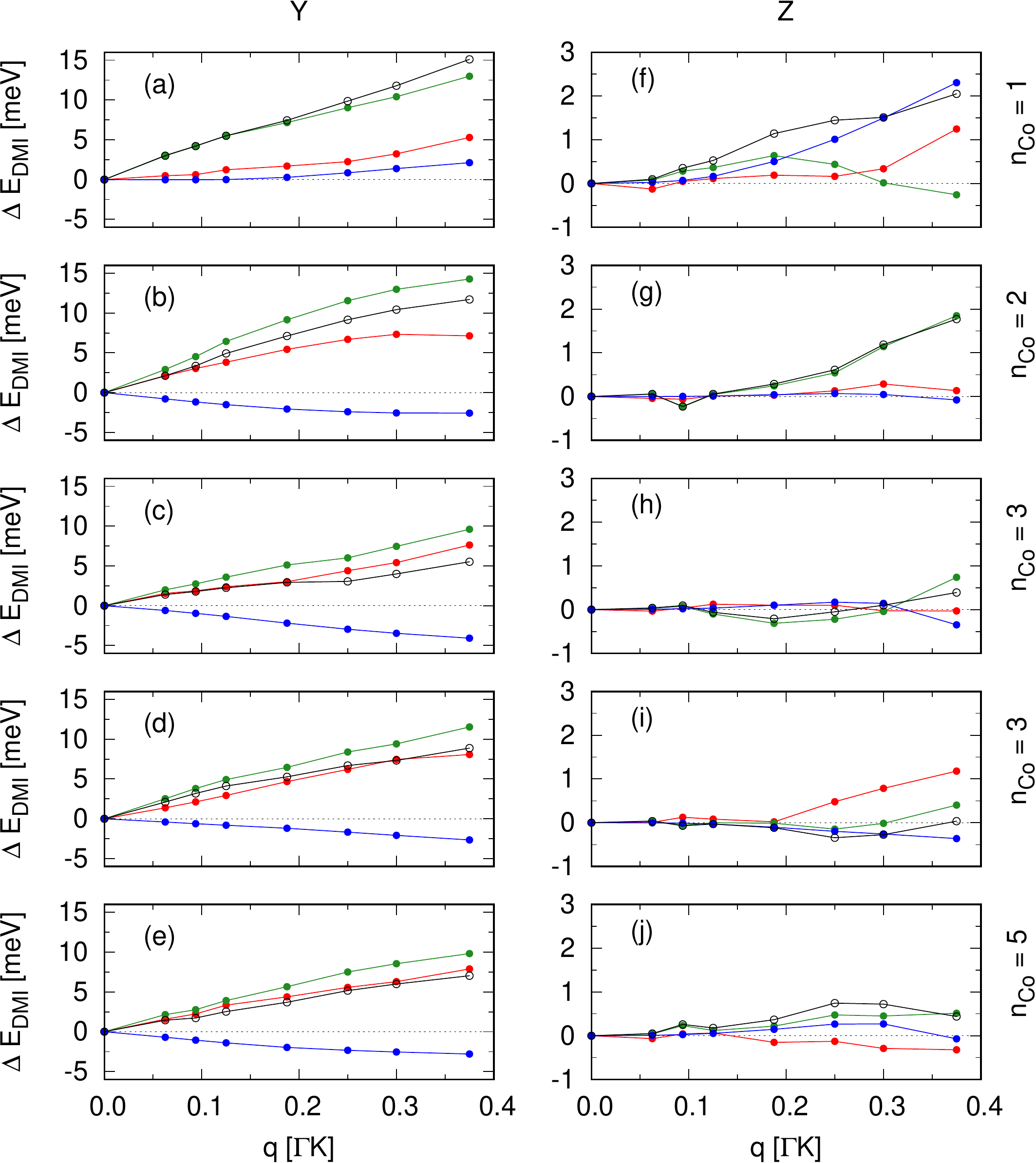}}
\caption{DMI energy $\Delta E_{DMI}^{\hat{s}_a}(\mathbf{q})$ for spin spirals
rotating in the plane perpendicular to  $Y$ (panel column a-e) and $Z$ (f-j)
axes and Co layer thicknesses (panel rows) in three different heterostructures:
Gr/Co$_n$/Pt$_5$ (red), Co$_n$/$Pt_5$ (green) and Gr/Co$_n$ (blue)
with $n=1-5$.
In black, the contribution sum from Co$_n$/Pt$_5 + $Gr/Co$_n$.
The spin spiral wavevector modulus $q$ is given in fractional units of the
distance $\Gamma K$ in the Brillouin zone (see Fig.~\ref{fig:sketch}b).}
\label{fig:partialslabs}
\end{figure}

At low $q$ values, the DMI energies for spins rotating in the plane 
perpendicular to $Y$ and $Z$ axes follow a nearly linear in $q$ and a $q^3$ 
dependence, respectively, as we describe next. 
These trends are observed in Fig.~\ref{fig:partialslabs} (red symbols and lines), 
which show $\Delta E_{DMI}^{\hat{s}_a}(\mathbf{q})$ 
for Gr/Co$_n$/Pt$_5$, $n=1-5$, slabs in the 
low-$q$ regime. Nevertheless, for the in-plane spins case the energies 
are too low to extract accurate quantitative results from a fit 
(note that the energies are one order of magnitude smaller in this 
spin geometry).
The evolution of $\Delta E_{DMI}^Y(\mathbf{q})$ with the Co layer thickness 
(red line) shows a significant magnitude variation, but no chirality change 
(i.e., no sign change). 
The $D_y$ effective parameter is extracted from the fit to a 2D 
model with nearest neighbour interactions in a hexagonal lattice 
with $C_{3v}$ symmetry, given by the expression 
\begin{align}
&\Delta E_{DMI}^{\hat s_a}(\mathbf{q}) = \nonumber \\
&4 S^2 \sin \theta \hat s_a \cdot \Big[ 0, D_y (1+2\cos \theta), 2D_z(\cos\theta-1) \Big] 
\label{eq:nn2d_model}
\end{align}
for flat spin spirals 
$\mathbf{q} = \frac{2\pi}{a} ( \frac{1}{N}, \frac{1}{N}, 0)$, 
where $\theta = \frac{2\pi}{N}$ are the corresponding angles between spins $S$ 
(assumed to have equal values).
This expression has a linear behaviour in the low-$q$ limit 
for spins rotating in the $XZ$ plane,
\begin{equation}
\Delta E_{DMI}^Y(\mathbf{q})  \approx 12 S^2 D_y \frac{2\pi}{N}
\label{eq:fitDy}
\end{equation}
We use this equation to fit the DFT energies, so that the resulting $D_y$ values 
are to be interpreted as \emph{effective}, since other contributions not 
considered in Eq.~\ref{eq:nn2d_model} also yield linear terms in $q$ for
$\hat s_a = Y$, such as inter-planar nearest neighbour bonds \cite{bib:crepieux98},  
second nearest and beyond neighbours.

\begin{figure}[tb!]
\centerline{\includegraphics[width=1\columnwidth]{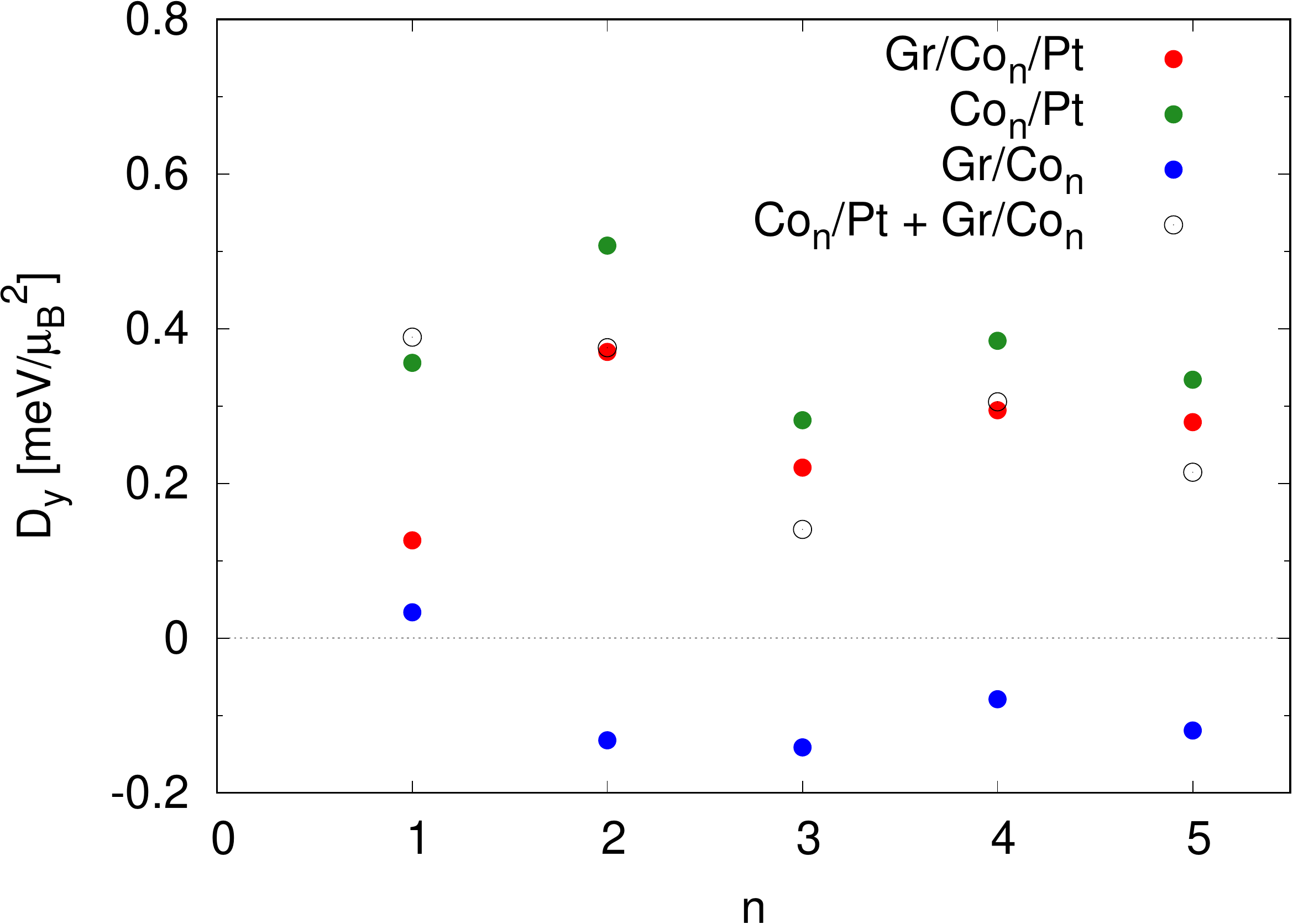}}
\caption{In-plane components of the effective $\mathbf{D}$ vector
obtained from fits of the data in the left-hand column of Fig.~\ref{fig:partialslabs}
to a linear law up to $q=0.25 |\Gamma K|$, where the averaged $m_S(\mathrm{Co})$
values over the Co layer have been used.
}
\label{fig:fits}
\end{figure}

To obtain $D_y$, we exclude the two highest $q$ points of each 
Fig.~\ref{fig:partialslabs} curve and use as spin moment value the 
average over the Co atomic layers. 
In the central Co atomic planes we find $m_S(\mathrm{Co}) \simeq 1.86$\,$\mu_B$. 
This value is slightly enhanced at the vacuum/Co and Co/Pt interfaces 
(1.94\,$\mu_B$ for $n_\mathrm{Co}=5$), while graphene has a demagnetizing 
effect (1.57\,$\mu_B$) (see also the Supplemental Material Fig.~\ref{SM-fig:cdd} bottom panels). 
Additionally, at the interfacial Pt atoms a spin polarization 
of 0.27\,$\mu_B$ is induced. This adds to $\Delta E_{DMI}^Y(\mathbf{q})$ a Co-Pt nearest 
neighbour contribution one order of magnitude smaller than that 
of the Co-Co DMI interactions. 
The resulting $D_y$ values, summarized in Fig.~\ref{fig:fits}, 
have a very good agreement with linear behaviour 
(the linear fit errors are $\le 4$\%).
The $D_y$ values oscillate for $n_{\mathrm{Co}}=1-3$ and 
converge to $D_y \simeq 0.3$\,meV$\mu_B^{-2}$ afterwards.
The fitted $D_y$ value for Co$_1$Pt$_5$ is in agreement 
with the literature calculated with GBT and 
flat spin spirals \cite{bib:belabbes16,bib:zimmermann19}, too,
which are close to 0.5\,meV$\mu_B^{-2}$. 
This methodology tends to yield larger energies than other 
electronic structure methods \cite{bib:zimmermann19}.

For spins rotating in the interface plane XY, the 
DMI energy is a third order effect in $q$,
\begin{equation}
\Delta E_{DMI}^Z(\mathbf{q}) \approx -4 S^2 D_z 
\Big( \frac{2\pi}{N} \Big)^3
\label{eq:fitDz}
\end{equation}
This means that at this geometry the interfaces sustain an 
effective $D$-vector with an out-of-plane component $D_z$ 
(note that a spin spiral with $q$-vector along $\Gamma M$
would not allow to resolve $D_z$), 
although the magnitude of this effect is small. 
The second column of Fig.~\ref{fig:partialslabs} shows that, indeed, 
the energies are an order of magnitude smaller than for 
spins rotating in the XZ plane. These data sets
do not allow for a good quality fit to a $q^3$ law, 
since the large $\xi(\mathrm{Pt})$ value 
magnify finite size effects. 
Slabs with a single Pt layer as substrate, which are shown 
in the Supplemental Material Fig.~\ref{SM-fig:partialslabs_pt1}, have a smoother behaviour.
In them $\Delta E_{DMI}^Z(\mathbf{q})$ changes its sign, 
i.e., alternating chirality of $D_z$,
as the Co layer grows beyond the monolayer thickness and tends 
toward small values when interfaces are decoupled at $n_\mathrm{Co}=3$.  
A sizable non-zero $D_z$ component can result in 
hybrid Bloch-N\'eel domain walls \cite{bib:lee09} 
and it has been postulated that it is responsible for an
asymmetric skyrmion Hall effect  \cite{bib:kim18}.
In addition,  the effect of the nearest-neighbour interplanar interaction 
energy term for $n_\mathrm{Co}=2$ is clearly distinguished in 
the $\Delta E_{DMI}^Y(\mathbf{q})$ curve. 

In the remainder of the paper we focus the analysis on the $D_y$ component only.
To ascertain whether the crossover at $n_{\mathrm{Co}}=3$ is correlated 
with the additivity of interfacial DMI, we have decomposed the 
heterostructure model slab into Co$_{1-5}$/Pt$_5$ and Gr/Co$_{1-5}$ 
partial slabs with the same atomic positions 
(green and blue lines and symbols, respectively, 
in Fig.~\ref{fig:partialslabs} and Fig.~\ref{fig:fits}). 
In the following, we call this form of analysis ``partial slab decomposition'' (PSD).
The sum of the corresponding $\Delta E_{DMI}^Y(\mathbf{q})$ and effective 
$D_y$ (black lines and open symbols) follows closely the 
Gr/Co$_{2-5}$/Pt$_5$ values, while there is a large difference for 
$n_{\mathrm{Co}}=1$. Although this would suggest that the Co/Pt and Gr/Co interfaces 
are already decoupled at $n_{\mathrm{Co}}=2$, their values still 
depend on the Co thickness. Hence, effective decoupling does not 
occur until at least $n_{\mathrm{Co}}=3$
In particular, for Gr/Co$_{2}$/Pt$_5$ the 
maximum value $D_y=0.5$\,meV$\mu_B^{-2}$ is obtained.

\begin{figure}[tb!]
\centerline{\includegraphics[width=1\columnwidth]{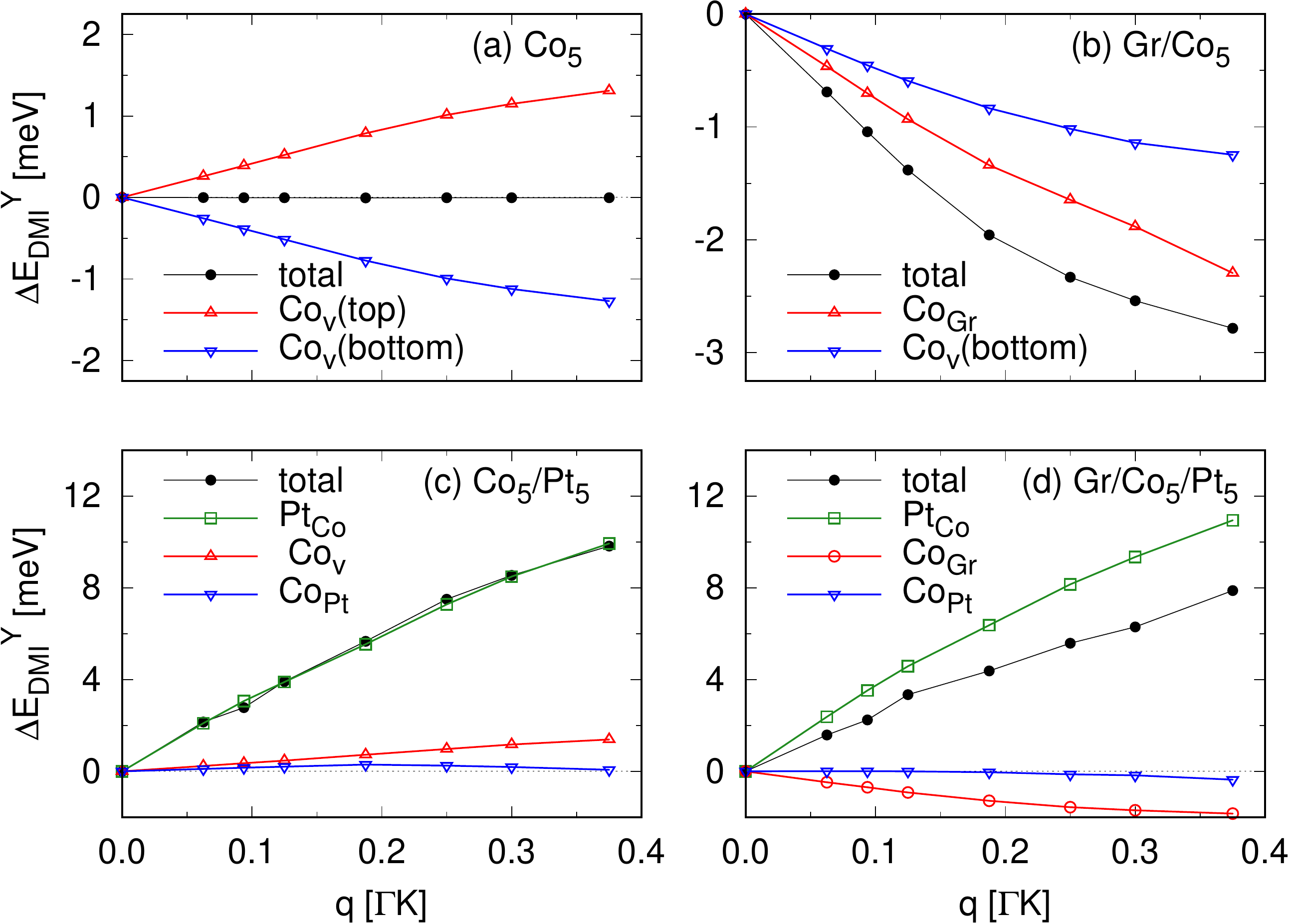}}
\caption{Contributions to the DMI energy $\Delta E_{DMI}^Y(\mathbf{q})$
(spins rotating in the plane perpendicular to $Y$ axis and
spiral wavevector along the $\Gamma K$ reciprocal direction)
from the individual interfacial atomic planes in Gr/Co$_5$/Pt$_5$
slab and its constituents.
The labels indicate the contributing atom and the subindices
indicate the interface the atom belongs to,  Co$_\mathrm{v}$
is the Co at the vaccum interface.  }
\label{fig:switch}
\end{figure}

The contribution of the Co/Pt interface accounts for most of the 
$\Delta E_{DMI}^Y(\mathbf{q})$ energy in the heterostructure. 
Importantly, at the Gr/Co interface the chirality is opposite (except in the GrCo 
bilayer) and of comparable magnitude to that of Co/Pt. 
This type of analysis, however, cannot determine if the effect of 
the Gr capping layer is to invert the chirality of the vacuum/Co interface.
For this, we have calculated $\Delta E_{DMI}^Y(\mathbf{q})$ with 
atomic SOC contributions selected {\it ad hoc}, a method we denote 
in the following with the acronym ASOD (atomic spin-orbit decomposition).
These results are shown in Fig.~\ref{fig:switch}.
This alternative method allows to assess the individual atom
contributions to interfacial DMI energy in a given system, since 
it is additive in the atomic SOC strength ($\xi$) by construction.
With ASOD we find that the 
contribution of the interfacial Pt atomic plane dominates 
the whole DMI effect, showing similar energies in both Gr/Co$_{5}$/Pt$_5$ and 
Co$_{5}$/Pt$_5$ slabs, whereas the Co plane in that same interface
has a negligible contribution compared to Co$_\mathrm{Gr}$, 
as also reported in the literature 
for similar $3d/5d$ interfaces~\cite{bib:yang15,bib:belabbes16}.   
This is explained in part by the strength of SOC at Pt, which is 
one order of magnitude larger than at Co ($\xi_\mathrm{Co}=74$\,meV and 
$\xi_\mathrm{Pt}=537$\,meV \cite{bib:blanco19}).
In the case of a bare Co$_5$ slab,
the two outermost atomic planes contribute with opposite chiralities
and sizable values, namely $|D_y| = 0.175$\,meV$\mu_B^{-2}$. 
Importantly, we find that the Co contribution changes from 
positive to negative $D_y$ and it is nearly doubled in magnitude 
when it is in contact with graphene in the Gr/Co$_5$ slab 
(the contribution of graphene itself is negligible due to the small 
$\xi_\mathrm{C}$ value). 
Therefore, graphene has the effect of reducing the 
net DMI of Gr/Co$_n$($fct$)/Pt heterostructures by inducing  
in the topmost Co plane an opposite chirality to that of the 
Co/Pt interface.
This has been observed by MOKE microscopy in domain wall propagation experiments,
although the graphene effect on the DMI was overestimated there 
to be about one half of that of the Co/Pt interface
(the reported spatial micromagnetic averages are 1.4 and -0.8\,mJ/m$^2$ 
for the Co/Pt~\cite{bib:luo19} and Gr/Co~\cite{bib:ajejas18} interfaces).

We have used two methods for the resolution of the DMI energy into 
its individual interfacial terms. 
The PSD method accounts for joint effect of 
(i) the SOC strength and (ii) the electronic structure modification 
at the interface when the constituents are brought together. 
In the ASOD method \emph{only the relativistic effect is being probed}. 
The latter method is less realistic, but more informative.
The importance of effect (ii) is manifested in the 
DMI tuning  by adsorption of light atoms. 
For example, H adsorption on Ni/Co/Pd/W induces chirality change \cite{bib:chen21}, 
with the advantage that H uptake and desorption is
a reversible process \cite{bib:chen21,bib:yang20}. 
DMI changes have also been characterized during oxidation of $3d/5d$ 
layered systems \cite{bib:belabbes16b,bib:chen20}. 
This DMI behaviour is associated to a charge density redistribution 
upon adsorption and, based on this mechanism, electrostatic 
properties such as surface dipoles, work-functions and 
electronegativity have been proposed as DMI 
descriptors \cite{bib:belabbes16b,bib:jia20}.

Table~\ref{tab:mtdipole} shows the perpendicular electric dipole $p_z$ 
of the interfacial C, Co and Pt atoms of the three partial slabs in 
the limit cases $n_\mathrm{Co} = 1$ and 5. 
It is evaluated as $p_z = -|e| \langle z \rangle_\mathrm{MT}$, where
the average position is evaluated as an integral over the charge density distribution
inside the muffin-tin. 
For the interfaces at the top of the slab,  $p_z>0 \quad (p_z<0)$ means that
the dipole points outward (inward), and viceversa for the
interfaces at the slab bottom.
For $n_\mathrm{Co}=1$, $p_z (\mathrm{Co})$ depends strongly on the interface type:
the dipole points inward when coated with graphene and outward when not. 
At the top of the $n_\mathrm{Co}=5$ slab a sign inversion of $p_z$ 
due to graphene is observed, too, alongside a reduction of the $p_z$ 
magnitude from 0.197 to 0.043\,a.u. (a factor 4.6). 
At the buried Co/Pt interface, Pt has also the effect of reducing the 
dipole of the interfacial Co$_\mathrm{Pt}$ atom by a factor 3.25 with respect to Co$_\mathrm{v}$, 
but no sign inversion occurs.
As reported in Ref.~\cite{bib:jia20}, the $p_z$ are correlated with DMI energies.
In the present case of the slabs with $n_\mathrm{Co}=5$, there is 
agreement between the signs of the $p_z$ of 
interfacial Co atoms (Table~\ref{tab:mtdipole}) and the signs of the 
contributions of these atoms to $\Delta E_{DMI}^Y$ (Fig.~\ref{fig:switch}).
However, there is no proportionality between the two magnitudes 
Note that dipoles and electronegativity are related to the $D$-vector 
by a non-linear analytical expression, which results in a linear correlation
between these properties for different adsorbed species \cite{bib:jia20}.

\begin{table}[tb!]
\caption{\label{tab:mtdipole}
Electric dipole (atomic units) in the direction perpendicular to the interfaces
inside the FLAPW muffin-tin spheres ($p_z$) for the interfacial atoms in the
Gr/Co$_n$/Pt$_5$, Co$_n$/Pt$_5$ and Gr/Co$_n$ slabs with $n=1,5$.
The subindex in the first column indicates the neighbouring
atomic layer in the interface.
At graphene, the $p_z$ of the two sublattice C atoms is averaged.
}
\begin{ruledtabular}
 \begin{tabular}{cccc}
atom   &  Gr/Co$_1$/Pt$_5$ & Co$_1$/Pt$_5$ &  Gr/Co$_1$ \\
\hline
C                & 0.014  &        & 0.014 \\
Co               & -0.111 &  0.153 & -0.267 \\
Pt$_\mathrm{Co}$ & -0.092 & -0.088 &   \\
\hline
atom   &  Gr/Co$_5$/Pt$_5$ & Co$_5$/Pt$_5$ &  Gr/Co$_5$ \\
\hline
C                                  & 0.013   &         & 0.013 \\
Co$_\mathrm{Gr}$ (Co$_\mathrm{v}$) &  -0.043 & (0.197) & -0.043  \\
Co$_\mathrm{Pt}$ (Co$_\mathrm{v}$) &  -0.059 & -0.060  & (-0.195) \\
Pt$_\mathrm{Co}$ &  -0.077 & -0.077 &     \\
 \end{tabular}
\end{ruledtabular}
\end{table}


We investigate next the influence on the DMI of the Co stacking, 
known to be a key factor to explain the magnetocrystalline anisotropy 
of Gr/Co/Pt heterostructures. 
When Co is pseudomorphically grown by 
intercalation in Gr/Pt(111), it results in a $fct$ stacking rather than 
$hcp$ \cite{bib:ajejas18,bib:ajejas20}, 
with stacking defects scattered throughout, nucleating 
predominantly near the Pt(111) substrate steps \cite{bib:blanco21}. 
In a perfect $fct$ heterostructure, DFT calculations show that 
the magnetocrystalline anisotropy energy (MAE) follows a      
bulk-like behaviour with in-plane anisotropy starting at $n_\mathrm{Co} = 8$
(0.09\,meV per Co atom, imposed by the lateral $fct$ lattice strain) 
and the critical thickness for 
perpendicular to in-plane switching at $n_\mathrm{Co}^c = 4$. 
However, the experimental switch occurs later, 
at $n_\mathrm{Co}^c \simeq 20$ (about 4\,nm), 
due to defects. The theory shows that each twin boundary 
contributes with nearly 1\,meV to out-of-plane anisotropy, 
as it introduces locally a $hcp$ stacked structure \cite{bib:blanco21}. 

\begin{figure}[tb!]
\centerline{\includegraphics[width=1\columnwidth]{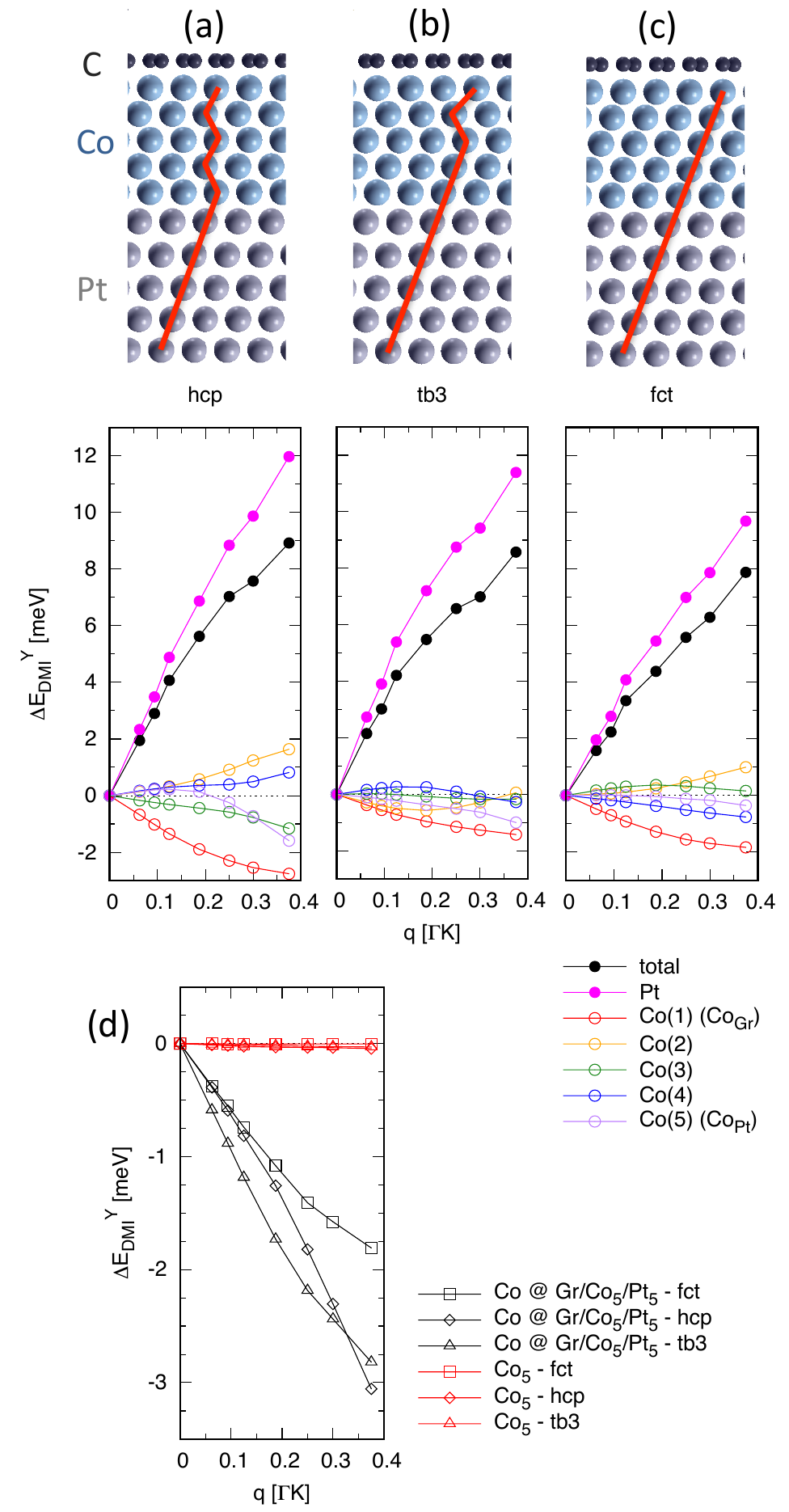}}
\caption{
(a-c) Top panels: cross section of the Gr/Co$_5$/Pt$_5$ slabs
with different Co stacking geometries:
perfect $hcp$, $fct$ with a twin boundary introduced in the
middle Co plane ($tb3$) and perfect $fct$.
A red line indicates the stacking sequences.
Bottom panels: for each geometry, 
total DMI energies ($\Delta E_{DMI}^Y(q)$) for spins rotating in the
plane perpendicular to $Y$ (black filled circles) and
those obtained when SOC is applied only to the indicated atomic planes
(coloured filled circles for Pt layer and empty for individual atomic
planes).
(d) Co layer contributions from Gr/Co$_5$/Pt$_5$ slabs (black symbols)
compared to bare Co$_5$ slabs with the three stacking types (red symbols). }
\label{fig:tb3}
\end{figure}

Fig.~\ref{fig:tb3} shows atom-resolved $\Delta E_{DMI}^Y(\mathbf{q})$ 
for Gr/Co$_5$/Pt$_5$ with Co in three different stackings (panels a-c): perfect $hcp$, 
$fct$ stacking with a twin boundary defect at the middle Co plane, 
labelled $tb3$ in the following, and perfect $fct$.
As in the original $fct$ heterostructure, 
with the $hcp$ and $tb3$ stackings the large $\xi_{\mathrm{Pt}}$ 
dominates the net DMI, contributing similar energies (black symbols)
In the case of a perfect $fct$ growth, the central region of the Co layer
is centrosymmetric, and therefore will not contribute to DMI. 
Nevertheless, in this finite-thickness model, we observe that the 
Co atomic plane contributions oscillate around zero. 
Oscillating values occur for the $tb3$ and the $hcp$ slabs, 
with larger energies in the latter. 
Note that for bare Co$_5$ slabs with the same geometries, 
i.e., in the absence of interfaces with Gr and Pt,
cancellations at the Co planes are almost total 
irrespective of the stacking type (red symbols in Fig.~\ref{fig:tb3}(d)). 
There is an overall negative contribution of the Co planes 
to the DMI (black symbols in Fig.~\ref{fig:tb3}(d))
that has its main origin at the Gr/Co interface.
Sizable DMI occurs at the Gr/Co and Co/Pt interfaces, 
while buried interfaces in Co that break inversion symmetry locally
contribute almost negligibly. 
This behaviour contrasts with that of the 
magnetic anisotropy, despite both properties sharing a common origin in the SOC,
with MAE being of $\xi^2$ order at this symmetry and DMI being linear in $\xi$.
Therefore, in Gr/Co/Pt heterostructure DMI and magnetocrystalline 
anisotropy will compete only close to the critical thicknesses, 
when the $fct$ bulk limit and the defect contributions compensate each other
to give a low MAE. 

\subsection{Reciprocal space analysis}
\label{sec:reciprocal}

So far, in the literature, interfacial DMI has been discussed 
in terms of two different mechanisms: a Rashba-like behaviour in Gr/Co 
triggered by the surface potential change $\nabla V$ induced by graphene 
adsorption \cite{bib:yang18} or, altenartively, a Fert-Levy-like behaviour in Co/Pt, 
where Pt SOC mediates the spin-flip of the Co itinerant electrons.
The aim of this section is to identify the nature of DMI at those 
interfaces based on information gathered at the reciprocal space.
The explicit dependence of DMI on each Bloch eigenstate $\epsilon_{nk}(\mathbf{q})$ 
is too intricate to be analyzed by bare eye,
due to the high density of dispersive bands in the Gr/Co$_n$/Pt$_5$ models 
(see Supplemental Material Fig.~\ref{SM-fig:bands_ptco1g}).
Note that each band is subject to lateral shifts due to the spin spirals 
and to degeneracy liftings, mainly at crossings between bands, due to SOC
(see Supplemental Material Fig.~\ref{SM-fig:bands_ptco1g}),
as shown by Sandratskii for the CoPt bilayer \cite{bib:sandratskii17}.
Instead, in our analysis we use quantities integrated in energies and in 
electron wavevectors $\mathbf{k}$.
 
\begin{figure}[tb!]
\centerline{\includegraphics[width=1\columnwidth]{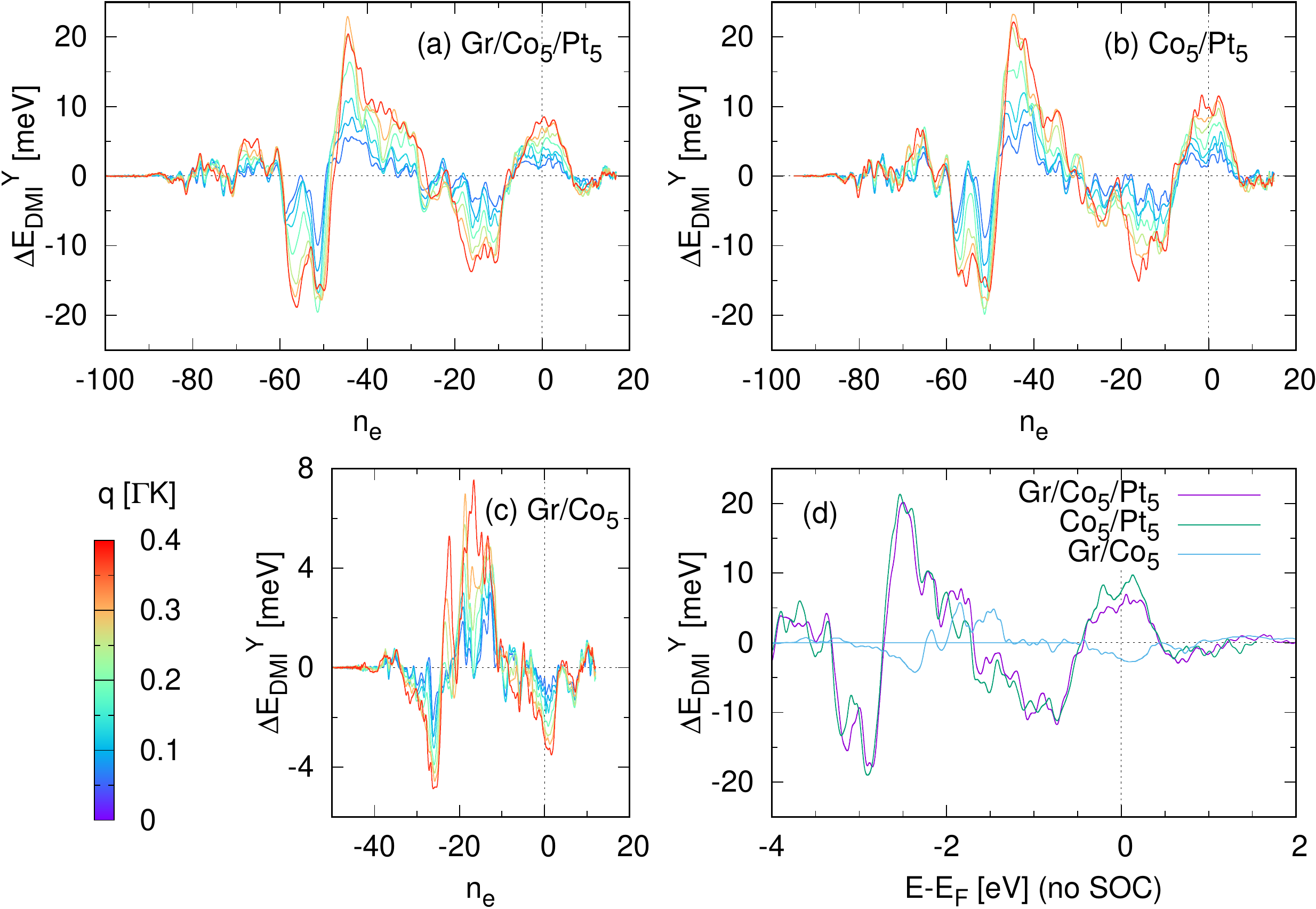}}
\caption{
(a-c) DMI energy ($\Delta E_{DMI}^Y(q)$) as a function of the band filling for
the Gr/Co$_5$/Pt$_5$, Co$_5$/Pt$_5$ and Gr/Co$_5$ slabs.
The color scale bar indicates the $q$ magnitude.
(d) The same curves plotted as a function of energy
(referred to the Fermi levels of each slab calculated without SOC)
for the particular $q$-point $q = 0.25|\Gamma K|$.
}
\label{fig:kelly}
\end{figure}

To analyze the spectral behaviour of the DMI chirality,
we plot the corresponding energies integrated in $\mathbf{k}$
as a function of the number of electrons $n_e$ for each spin spiral vector $\mathbf{q}$.
This is similar to the MAEs in the force theorem approach \cite{bib:daalderop94,bib:blanco19}:
\begin{align}
\Delta E_{DMI}^Y(n_e;\mathbf{q}) =  \sum_{nk} &
    \epsilon_{nk}^{Y^+}(\mathbf{q}) 
f \Big( \epsilon_{nk}^{Y^+}(\mathbf{q})-\epsilon_F^{Y^+}(n_e;\mathbf{q}) \Big) \nonumber\\
& - \epsilon_{nk}^{Y^-}(\mathbf{q}) 
f \Big( \epsilon_{nk}^{Y^-}(\mathbf{q})-\epsilon_F^{Y^-}(n_e;\mathbf{q}) \Big)
\label{eq:kelly}
\end{align}
where $\mathbf{q}$ is a spin spiral wavevector along $\Gamma K$, 
the sum runs over the eigenvalues $\epsilon_{nk}$
calculated for opposite spin rotation axes ($Y^+$ and $Y^-$)
and the Fermi levels correspond to the filling up with $n_e$ electrons of the bands 
of each individual calculation with $\mathbf{q}$ and $Y^+$ or $Y^-$.
The  $\Delta E_{DMI}^Y(n_e;\mathbf{q})$ curves for Gr/Co$_5$/Pt$_5$, Co$_5$/Pt$_5$
and Gr/Co$_5$ slabs are qualitatively invariant with $q$
(see  Fig.~\ref{fig:kelly}):
the nodes in the curves are almost invariant and peaks only
increase in amplitude with $q$, resulting in the linear
dependence observed in the previous section. This behaviour occurs
also in the less complex CoPt bilayer, where it has been explained \cite{bib:sandratskii17}
by the hybridization  by the spiral of electron states 
with wavevectors $\mathbf{k}$ and $\mathbf{k} \pm \mathbf{q}$, 
which have similar character regardless of $\mathbf{q}$, giving rise
to the dependence of Eq.~\ref{eq:nn2d_model} 
\footnote{
Some selection rules are found concerning 
the orbital character $|l,m \rangle$ of the states with 
wavevectors $\mathbf{k}$ and $\mathbf{k} \pm \mathbf{q}$, 
When the $\mathbf{q}$-dependence is perturbatively considered 
in the analytical $D$-vector expression, 
a dipolar term appears that couples different $l$-channels,
as shown in Ref.~\cite{bib:jia20}.
If only the inversion symmetry breaking is considered,
as in Ref.~\cite{bib:yang20}, only the couplings 
between $|2,m\rangle$ orbitals of suitable symmetry 
(determined by the SOC term in the hamiltonian \cite{bib:abate65}) survive.}.
As observed in Fig.~\ref{fig:kelly}, this scenario is not affected 
by the presence of the large number of 
additional bands of a thicker Pt substrate, which is the reciprocal space 
confirmation that the DMI is localized within a very short spatial range 
(bondlength distance) of the interface. 
At neutrality ($n_e=0$) the known positive $\Delta E_{DMI}^Y(0;\mathbf{q})$ values for
Gr/Co$_5$/Pt$_5$ and Co$_5$/Pt$_5$, and negative for Gr/Co$_5$ are retrieved.
The sign of the DMI energy is kept under small variations of $n_e$ around the
neutrality condition.
When represented as a function of binding energies,
$\Delta E_{DMI}^Y(n_e;\mathbf{q})$ has the last node before neutrality at the
filling corresponding to $E_F-0.5$\,eV in the three slabs, as shown in
the Fig.~\ref{fig:kelly}(d) panel for $\mathbf{q}=\frac{2\pi}{a}(\frac{1}{12},\frac{1}{12},0)$.
This means that integration of eigenenergies in this energy window of 0.5\,eV 
would \emph{effectively} reproduce the net DMI, albeit non-zero DMI 
contributions occur throughout the whole available energy spectrum.
This is evident with a Pt substrate, since the $5d-3d$ hybrid bandwidth
spans several eV. However, Fig.~\ref{fig:kelly}(c,d) shows the same 
qualitative behaviour for Gr/Co$_5$, pointing out that there is a 
similar DMI mechanism here due only to the Co SOC contributions.
Note that the 0.5\,eV window coincidence 
for the three slabs is fortuitous, as it depends on the particular 
details of each band structure.

\begin{figure}[tb!]
\centerline{\includegraphics[width=1\columnwidth]{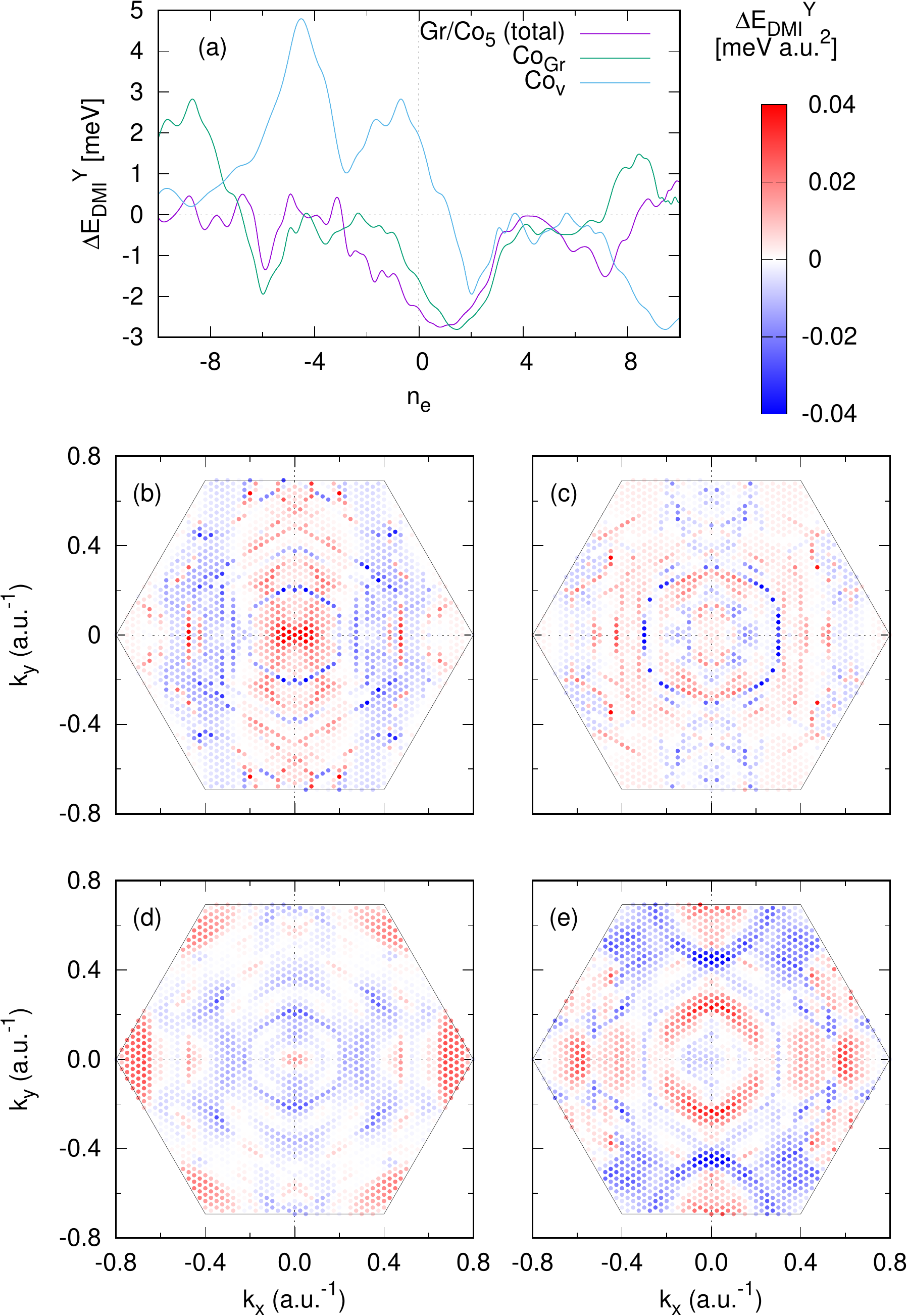}}
\caption{
(a) $\Delta E_{DMI}(n_e)$ at $q = 0.25|\Gamma K|$ for the
Gr/Co$_5$ slab (purple) compared with the curves calculated for SOC
applied only to the Co$_{\mathrm{Gr}}$ plane of that slab (green) and to
Co$_{\mathrm{v}}$ of the clean Co$_5$ slab (blue)
(note that the latter is centrosymmetric, thus the
total $\Delta E_{DMI}^Y$ in the slab amounts to zero).
(b-e) For the same $\mathbf{q}$-vector,
$\Delta E_{DMI}^Y$ at $n_e=0$ resolved in $k$-space.
Panels (b,d) correspond to the Co$_{Gr}$ interfacial atomic plane and
(c,e) to Co$_{v}$ (Co/vaccum interface).
Data of panels (b,c) are calculated by integration over all the occupied bands.
In panels (d,e), the integral energy range is restricted to
a stripe of 0.5\,eV below the Fermi level.}
\label{fig:1bz}
\end{figure}

With focus on the DMI chirality inversion of the vacuum/Co interface 
upon capping with graphene, we first verify that the 
$\Delta E_{DMI}(0;\mathbf{q})$ values reproduce the sign change when evaluated with 
only the corresponding interfacial individual atomic SOC strength 
$\xi_\mathrm{Co}$ (see Fig.~\ref{fig:1bz}(a)).
We now turn to a $\mathbf{k}$-resolved analysis of these quantities. 
Fig.~\ref{fig:1bz}(b,c) show that DMI is not localized in the $\mathbf{k}$-space,
but broad regions of the Brillouin zone (BZ) contribute with opposite chiralities
to the final net $\Delta E_{DMI} (0;\mathbf{q})$ in both vacuum/Co and Gr/Co interfaces. 
Owing to the found effective energy window, in Fig.~\ref{fig:1bz}(d,e) we 
restrict the $\mathbf{k}$-resolved analysis to an integral over states 
within a window of 0.5\,eV below the Fermi level. We observe that the Co-C hybrid bands,
with nearly conical dispersion at the $K,K'$ special points yield a positive contribution 
to $\Delta E_{DMI}(0;\mathbf{q})$ (red spots at the BZ vertices), 
whereas the rest of the BZ contributes with negative values. 
In other words, the distinctive feature of the graphene adsorption 
on the bandstructure, namely the conical band
of mainly Co-$d_{z^2}$ character and also small weight in $d_{xz,yz}$ orbitals,
actually contributes to a chirality opposite to the
observed one. The conclusion is that the DMI of the Gr/Co interface 
cannot be attributed to individual Co-C hybrid bands near the Fermi level.
For this reason, a model where the $D$ vector is estimated from 
a Rashba hamiltonian $\alpha_R (\mathbf{\sigma} \times \mathbf{k})_z$
of a single band in a ferromagnetically coupled 
environment \cite{bib:kim13,bib:kundu15,bib:yang18,bib:hallal21} 
is not suitable for the Gr/Co interface. 
Instead, the contributing Co-C interactions extend to the whole spectrum.
On this basis we can state that the nature of interfacial DMI at Gr/Co and Co/Pt is the same,
namely strongly hybridized SOC-split $d$-bands.
We recall that Rashba band splitting requires not only a $\nabla V$, 
which can be indeed enhanced upon adsorption if this increases the 
asymmetry of the charge distribution at the surface \cite{bib:krupin05},
but also a substantial SOC strength $\xi$ \cite{bib:lashell96,bib:petersen00}.

\section{Conclusions}
\label{sec:conclusions}

Our interface-resolved DFT study of DMI in Gr/Co$_n$/Pt(111)
heterostructures with varying Co layer thickness shows that the 
regime of additivity of interfacial DMI is reached already 
at $n=3$ atomic planes and also that the $D$-vectors have an almost negligible out-of-plane 
component. As the perpendicular magnetocrystalline anisotropy 
prevails at much larger $n$ values \cite{bib:blanco21},
a sizable DMI interaction to have spin canting and 
chiral exchange effects are expected to be robust. 
However, unlike the magnetocrystalline anisotropy itself, interfacial DMI is insensitive 
to the internal structure of the Co layer.
The observed DMI in domain wall propagation suggests a
comparable interface DMI strength but opposite sign at Gr/Co and 
Co/Pt interfaces \cite{bib:blanco21}. This is confirmed by our calculations. 
Indeed, we find that the graphene layer has the effect of 
inverting the chirality of the vacuum/Co interface.

The Gr/Co and Co/Pt interfacial DMI has been classified 
as being of different nature, namely, Rashba and Fert-Levy mechanisms at 
Gr/Co and Co/Pt, respectively. 
Our study leads to the conclusion that this classification is subjective.
Those models correspond to two limiting cases of the same physics, 
as illustrated by the Gr/Co$_n$/Pt system.
The electrostatic dipole (a magnitude identified as a DMI descriptor)
at the vacuum/Co surface is reversed and increases in 
magnitude upon capping with graphene. 
Nevertheless, this does not mean that a Rashba SOC term is induced: 
a spectral resolution of the DMI energy of Gr/Co reveals that it is 
not linked to an individual band splitting. Instead, 
it results from the superposition of many hybridized bands with weight on the 
Co atoms, similarly to the other Co/Pt interface.

\begin{acknowledgments}
%
Pojects RTI2018-097895-B-C41 and PID2019-103910GB-I00, 
funded by MCIN/AEI/10.13039/501100011033/ and FEDER {\it Una manera de hacer Europa};
GIU18/138 by Universidad del Pa\'{\i}s Vasco UPV/EHU;
IT-1246-19 and IT-1260-19 by Gobierno Vasco.
Computational resources were provided by the DIPC computing center.
\end{acknowledgments}

\newpage

\bibliography{dmi}

\begin{thebibliography}{54}%
\makeatletter
\providecommand \@ifxundefined [1]{%
 \@ifx{#1\undefined}
}%
\providecommand \@ifnum [1]{%
 \ifnum #1\expandafter \@firstoftwo
 \else \expandafter \@secondoftwo
 \fi
}%
\providecommand \@ifx [1]{%
 \ifx #1\expandafter \@firstoftwo
 \else \expandafter \@secondoftwo
 \fi
}%
\providecommand \natexlab [1]{#1}%
\providecommand \enquote  [1]{``#1''}%
\providecommand \bibnamefont  [1]{#1}%
\providecommand \bibfnamefont [1]{#1}%
\providecommand \citenamefont [1]{#1}%
\providecommand \href@noop [0]{\@secondoftwo}%
\providecommand \href [0]{\begingroup \@sanitize@url \@href}%
\providecommand \@href[1]{\@@startlink{#1}\@@href}%
\providecommand \@@href[1]{\endgroup#1\@@endlink}%
\providecommand \@sanitize@url [0]{\catcode `\\12\catcode `\$12\catcode
  `\&12\catcode `\#12\catcode `\^12\catcode `\_12\catcode `\%12\relax}%
\providecommand \@@startlink[1]{}%
\providecommand \@@endlink[0]{}%
\providecommand \url  [0]{\begingroup\@sanitize@url \@url }%
\providecommand \@url [1]{\endgroup\@href {#1}{\urlprefix }}%
\providecommand \urlprefix  [0]{URL }%
\providecommand \Eprint [0]{\href }%
\providecommand \doibase [0]{https://doi.org/}%
\providecommand \selectlanguage [0]{\@gobble}%
\providecommand \bibinfo  [0]{\@secondoftwo}%
\providecommand \bibfield  [0]{\@secondoftwo}%
\providecommand \translation [1]{[#1]}%
\providecommand \BibitemOpen [0]{}%
\providecommand \bibitemStop [0]{}%
\providecommand \bibitemNoStop [0]{.\EOS\space}%
\providecommand \EOS [0]{\spacefactor3000\relax}%
\providecommand \BibitemShut  [1]{\csname bibitem#1\endcsname}%
\let\auto@bib@innerbib\@empty
\bibitem [{\citenamefont {Dzialoshinskii}(1957)}]{bib:dzyaloshinskii57}%
  \BibitemOpen
  \bibfield  {author} {\bibinfo {author} {\bibfnamefont {I.~E.}\ \bibnamefont
  {Dzialoshinskii}},\ }\href@noop {} {\bibfield  {journal} {\bibinfo  {journal}
  {Soviet Phys. JETP}\ }\textbf {\bibinfo {volume} {5}},\ \bibinfo {pages}
  {1259} (\bibinfo {year} {1957})}\BibitemShut {NoStop}%
\bibitem [{\citenamefont {Moriya}(1960{\natexlab{a}})}]{bib:moriya60}%
  \BibitemOpen
  \bibfield  {author} {\bibinfo {author} {\bibfnamefont {T.}~\bibnamefont
  {Moriya}},\ }\href {https://doi.org/10.1103/PhysRevLett.4.228} {\bibfield
  {journal} {\bibinfo  {journal} {Phys. Rev. Lett.}\ }\textbf {\bibinfo
  {volume} {4}},\ \bibinfo {pages} {228} (\bibinfo {year}
  {1960}{\natexlab{a}})}\BibitemShut {NoStop}%
\bibitem [{\citenamefont {Moriya}(1960{\natexlab{b}})}]{bib:moriya60b}%
  \BibitemOpen
  \bibfield  {author} {\bibinfo {author} {\bibfnamefont {T.}~\bibnamefont
  {Moriya}},\ }\href {https://doi.org/10.1103/PhysRev.120.91} {\bibfield
  {journal} {\bibinfo  {journal} {Phys. Rev.}\ }\textbf {\bibinfo {volume}
  {120}},\ \bibinfo {pages} {91} (\bibinfo {year}
  {1960}{\natexlab{b}})}\BibitemShut {NoStop}%
\bibitem [{\citenamefont {Fert}\ and\ \citenamefont {Levy}(1980)}]{bib:fert80}%
  \BibitemOpen
  \bibfield  {author} {\bibinfo {author} {\bibfnamefont {A.}~\bibnamefont
  {Fert}}\ and\ \bibinfo {author} {\bibfnamefont {P.~M.}\ \bibnamefont
  {Levy}},\ }\href {https://doi.org/10.1103/PhysRevLett.44.1538} {\bibfield
  {journal} {\bibinfo  {journal} {Phys. Rev. Lett.}\ }\textbf {\bibinfo
  {volume} {44}},\ \bibinfo {pages} {1538} (\bibinfo {year}
  {1980})}\BibitemShut {NoStop}%
\bibitem [{\citenamefont {Levy}\ and\ \citenamefont {Fert}(1981)}]{bib:fert81}%
  \BibitemOpen
  \bibfield  {author} {\bibinfo {author} {\bibfnamefont {P.~M.}\ \bibnamefont
  {Levy}}\ and\ \bibinfo {author} {\bibfnamefont {A.}~\bibnamefont {Fert}},\
  }\href {https://doi.org/10.1103/PhysRevB.23.4667} {\bibfield  {journal}
  {\bibinfo  {journal} {Phys. Rev. B}\ }\textbf {\bibinfo {volume} {23}},\
  \bibinfo {pages} {4667} (\bibinfo {year} {1981})}\BibitemShut {NoStop}%
\bibitem [{\citenamefont {Bode}\ \emph {et~al.}(2007)\citenamefont {Bode},
  \citenamefont {Heide}, \citenamefont {von Bergmann}, \citenamefont
  {Ferriani}, \citenamefont {Heinze}, \citenamefont {Bihlmayer}, \citenamefont
  {Kubetzka}, \citenamefont {Pietzsch}, \citenamefont {Bl{\"u}gel},\ and\
  \citenamefont {Wiesendanger}}]{bib:bode07}%
  \BibitemOpen
  \bibfield  {author} {\bibinfo {author} {\bibfnamefont {M.}~\bibnamefont
  {Bode}}, \bibinfo {author} {\bibfnamefont {M.}~\bibnamefont {Heide}},
  \bibinfo {author} {\bibfnamefont {K.}~\bibnamefont {von Bergmann}}, \bibinfo
  {author} {\bibfnamefont {P.}~\bibnamefont {Ferriani}}, \bibinfo {author}
  {\bibfnamefont {S.}~\bibnamefont {Heinze}}, \bibinfo {author} {\bibfnamefont
  {G.}~\bibnamefont {Bihlmayer}}, \bibinfo {author} {\bibfnamefont
  {A.}~\bibnamefont {Kubetzka}}, \bibinfo {author} {\bibfnamefont
  {O.}~\bibnamefont {Pietzsch}}, \bibinfo {author} {\bibfnamefont
  {S.}~\bibnamefont {Bl{\"u}gel}},\ and\ \bibinfo {author} {\bibfnamefont
  {R.}~\bibnamefont {Wiesendanger}},\ }\href
  {https://doi.org/10.1038/nature05802} {\bibfield  {journal} {\bibinfo
  {journal} {Nature}\ }\textbf {\bibinfo {volume} {447}},\ \bibinfo {pages}
  {190} (\bibinfo {year} {2007})}\BibitemShut {NoStop}%
\bibitem [{\citenamefont {Schmitt}\ \emph {et~al.}(2019)\citenamefont
  {Schmitt}, \citenamefont {Moras}, \citenamefont {Bihlmayer}, \citenamefont
  {Cotsakis}, \citenamefont {Vogt}, \citenamefont {Kemmer}, \citenamefont
  {Belabbes}, \citenamefont {Sheverdyaeva}, \citenamefont {Kundu},
  \citenamefont {Carbone}, \citenamefont {Bl{\"u}gel},\ and\ \citenamefont
  {Bode}}]{bib:schmitt19}%
  \BibitemOpen
  \bibfield  {author} {\bibinfo {author} {\bibfnamefont {M.}~\bibnamefont
  {Schmitt}}, \bibinfo {author} {\bibfnamefont {P.}~\bibnamefont {Moras}},
  \bibinfo {author} {\bibfnamefont {G.}~\bibnamefont {Bihlmayer}}, \bibinfo
  {author} {\bibfnamefont {R.}~\bibnamefont {Cotsakis}}, \bibinfo {author}
  {\bibfnamefont {M.}~\bibnamefont {Vogt}}, \bibinfo {author} {\bibfnamefont
  {J.}~\bibnamefont {Kemmer}}, \bibinfo {author} {\bibfnamefont
  {A.}~\bibnamefont {Belabbes}}, \bibinfo {author} {\bibfnamefont {P.~M.}\
  \bibnamefont {Sheverdyaeva}}, \bibinfo {author} {\bibfnamefont {A.~K.}\
  \bibnamefont {Kundu}}, \bibinfo {author} {\bibfnamefont {C.}~\bibnamefont
  {Carbone}}, \bibinfo {author} {\bibfnamefont {S.}~\bibnamefont
  {Bl{\"u}gel}},\ and\ \bibinfo {author} {\bibfnamefont {M.}~\bibnamefont
  {Bode}},\ }\href {https://doi.org/10.1038/s41467-019-10515-3} {\bibfield
  {journal} {\bibinfo  {journal} {Nature Communications}\ }\textbf {\bibinfo
  {volume} {10}},\ \bibinfo {pages} {2610} (\bibinfo {year}
  {2019})}\BibitemShut {NoStop}%
\bibitem [{\citenamefont {Wiesendanger}(2016)}]{bib:wiesendanger16}%
  \BibitemOpen
  \bibfield  {author} {\bibinfo {author} {\bibfnamefont {R.}~\bibnamefont
  {Wiesendanger}},\ }\href {https://doi.org/10.1038/natrevmats.2016.44}
  {\bibfield  {journal} {\bibinfo  {journal} {Nature Reviews Materials}\
  }\textbf {\bibinfo {volume} {1}},\ \bibinfo {pages} {16044} (\bibinfo {year}
  {2016})}\BibitemShut {NoStop}%
\bibitem [{\citenamefont {Fert}\ \emph {et~al.}(2017)\citenamefont {Fert},
  \citenamefont {Reyren},\ and\ \citenamefont {Cros}}]{bib:fert17}%
  \BibitemOpen
  \bibfield  {author} {\bibinfo {author} {\bibfnamefont {A.}~\bibnamefont
  {Fert}}, \bibinfo {author} {\bibfnamefont {N.}~\bibnamefont {Reyren}},\ and\
  \bibinfo {author} {\bibfnamefont {V.}~\bibnamefont {Cros}},\ }\href
  {https://doi.org/10.1038/natrevmats.2017.31} {\bibfield  {journal} {\bibinfo
  {journal} {Nature Reviews Materials}\ }\textbf {\bibinfo {volume} {2}},\
  \bibinfo {pages} {17031} (\bibinfo {year} {2017})}\BibitemShut {NoStop}%
\bibitem [{\citenamefont {Bogdanov}\ and\ \citenamefont
  {Panagopoulos}(2020)}]{bib:bogdanov20}%
  \BibitemOpen
  \bibfield  {author} {\bibinfo {author} {\bibfnamefont {A.~N.}\ \bibnamefont
  {Bogdanov}}\ and\ \bibinfo {author} {\bibfnamefont {C.}~\bibnamefont
  {Panagopoulos}},\ }\href {https://doi.org/10.1038/s42254-020-0203-7}
  {\bibfield  {journal} {\bibinfo  {journal} {Nature Reviews Physics}\ }\textbf
  {\bibinfo {volume} {2}},\ \bibinfo {pages} {492} (\bibinfo {year}
  {2020})}\BibitemShut {NoStop}%
\bibitem [{\citenamefont {Heide}\ \emph {et~al.}(2008)\citenamefont {Heide},
  \citenamefont {Bihlmayer},\ and\ \citenamefont {Bl\"ugel}}]{bib:heide08}%
  \BibitemOpen
  \bibfield  {author} {\bibinfo {author} {\bibfnamefont {M.}~\bibnamefont
  {Heide}}, \bibinfo {author} {\bibfnamefont {G.}~\bibnamefont {Bihlmayer}},\
  and\ \bibinfo {author} {\bibfnamefont {S.}~\bibnamefont {Bl\"ugel}},\ }\href
  {https://doi.org/10.1103/PhysRevB.78.140403} {\bibfield  {journal} {\bibinfo
  {journal} {Phys. Rev. B}\ }\textbf {\bibinfo {volume} {78}},\ \bibinfo
  {pages} {140403} (\bibinfo {year} {2008})}\BibitemShut {NoStop}%
\bibitem [{\citenamefont {Luo}\ \emph {et~al.}(2019)\citenamefont {Luo},
  \citenamefont {Dao}, \citenamefont {Hrabec}, \citenamefont {Vijayakumar},
  \citenamefont {Kleibert}, \citenamefont {Baumgartner}, \citenamefont {Kirk},
  \citenamefont {Cui}, \citenamefont {Savchenko}, \citenamefont {Krishnaswamy},
  \citenamefont {Heyderman},\ and\ \citenamefont {Gambardella}}]{bib:luo19}%
  \BibitemOpen
  \bibfield  {author} {\bibinfo {author} {\bibfnamefont {Z.}~\bibnamefont
  {Luo}}, \bibinfo {author} {\bibfnamefont {T.~P.}\ \bibnamefont {Dao}},
  \bibinfo {author} {\bibfnamefont {A.}~\bibnamefont {Hrabec}}, \bibinfo
  {author} {\bibfnamefont {J.}~\bibnamefont {Vijayakumar}}, \bibinfo {author}
  {\bibfnamefont {A.}~\bibnamefont {Kleibert}}, \bibinfo {author}
  {\bibfnamefont {M.}~\bibnamefont {Baumgartner}}, \bibinfo {author}
  {\bibfnamefont {E.}~\bibnamefont {Kirk}}, \bibinfo {author} {\bibfnamefont
  {J.}~\bibnamefont {Cui}}, \bibinfo {author} {\bibfnamefont {T.}~\bibnamefont
  {Savchenko}}, \bibinfo {author} {\bibfnamefont {G.}~\bibnamefont
  {Krishnaswamy}}, \bibinfo {author} {\bibfnamefont {L.~J.}\ \bibnamefont
  {Heyderman}},\ and\ \bibinfo {author} {\bibfnamefont {P.}~\bibnamefont
  {Gambardella}},\ }\href {https://doi.org/10.1126/science.aau7913} {\bibfield
  {journal} {\bibinfo  {journal} {Science}\ }\textbf {\bibinfo {volume}
  {363}},\ \bibinfo {pages} {1435} (\bibinfo {year} {2019})},\ \Eprint
  {https://arxiv.org/abs/https://science.sciencemag.org/content/363/6434/1435.full.pdf}
  {https://science.sciencemag.org/content/363/6434/1435.full.pdf} \BibitemShut
  {NoStop}%
\bibitem [{\citenamefont {Hrabec}\ \emph {et~al.}(2020)\citenamefont {Hrabec},
  \citenamefont {Luo}, \citenamefont {Heyderman},\ and\ \citenamefont
  {Gambardella}}]{bib:hrabec20}%
  \BibitemOpen
  \bibfield  {author} {\bibinfo {author} {\bibfnamefont {A.}~\bibnamefont
  {Hrabec}}, \bibinfo {author} {\bibfnamefont {Z.}~\bibnamefont {Luo}},
  \bibinfo {author} {\bibfnamefont {L.~J.}\ \bibnamefont {Heyderman}},\ and\
  \bibinfo {author} {\bibfnamefont {P.}~\bibnamefont {Gambardella}},\ }\href
  {https://doi.org/10.1063/5.0021184} {\bibfield  {journal} {\bibinfo
  {journal} {Applied Physics Letters}\ }\textbf {\bibinfo {volume} {117}},\
  \bibinfo {pages} {130503} (\bibinfo {year} {2020})}\BibitemShut {NoStop}%
\bibitem [{\citenamefont {Vedmedenko}\ \emph {et~al.}(2020)\citenamefont
  {Vedmedenko}, \citenamefont {Kawakami}, \citenamefont {Sheka}, \citenamefont
  {Gambardella}, \citenamefont {Kirilyuk}, \citenamefont {Hirohata},
  \citenamefont {Binek}, \citenamefont {Chubykalo-Fesenko}, \citenamefont
  {Sanvito}, \citenamefont {Kirby}, \citenamefont {Grollier}, \citenamefont
  {Everschor-Sitte}, \citenamefont {Kampfrath}, \citenamefont {You},\ and\
  \citenamefont {Berger}}]{bib:roadmap20}%
  \BibitemOpen
  \bibfield  {author} {\bibinfo {author} {\bibfnamefont {E.~Y.}\ \bibnamefont
  {Vedmedenko}}, \bibinfo {author} {\bibfnamefont {R.~K.}\ \bibnamefont
  {Kawakami}}, \bibinfo {author} {\bibfnamefont {D.~D.}\ \bibnamefont {Sheka}},
  \bibinfo {author} {\bibfnamefont {P.}~\bibnamefont {Gambardella}}, \bibinfo
  {author} {\bibfnamefont {A.}~\bibnamefont {Kirilyuk}}, \bibinfo {author}
  {\bibfnamefont {A.}~\bibnamefont {Hirohata}}, \bibinfo {author}
  {\bibfnamefont {C.}~\bibnamefont {Binek}}, \bibinfo {author} {\bibfnamefont
  {O.}~\bibnamefont {Chubykalo-Fesenko}}, \bibinfo {author} {\bibfnamefont
  {S.}~\bibnamefont {Sanvito}}, \bibinfo {author} {\bibfnamefont {B.~J.}\
  \bibnamefont {Kirby}}, \bibinfo {author} {\bibfnamefont {J.}~\bibnamefont
  {Grollier}}, \bibinfo {author} {\bibfnamefont {K.}~\bibnamefont
  {Everschor-Sitte}}, \bibinfo {author} {\bibfnamefont {T.}~\bibnamefont
  {Kampfrath}}, \bibinfo {author} {\bibfnamefont {C.-Y.}\ \bibnamefont {You}},\
  and\ \bibinfo {author} {\bibfnamefont {A.}~\bibnamefont {Berger}},\ }\href
  {https://doi.org/10.1088/1361-6463/ab9d98} {\bibfield  {journal} {\bibinfo
  {journal} {Journal of Physics D: Applied Physics}\ }\textbf {\bibinfo
  {volume} {53}},\ \bibinfo {pages} {453001} (\bibinfo {year}
  {2020})}\BibitemShut {NoStop}%
\bibitem [{\citenamefont {Moreau-Luchaire}\ \emph {et~al.}(2016)\citenamefont
  {Moreau-Luchaire}, \citenamefont {Moutafis}, \citenamefont {Reyren},
  \citenamefont {Sampaio}, \citenamefont {Vaz}, \citenamefont {Van~Horne},
  \citenamefont {Bouzehouane}, \citenamefont {Garcia}, \citenamefont
  {Deranlot}, \citenamefont {Warnicke}, \citenamefont {Wohlh{\"u}ter},
  \citenamefont {George}, \citenamefont {Weigand}, \citenamefont {Raabe},
  \citenamefont {Cros},\ and\ \citenamefont {Fert}}]{bib:moreau16}%
  \BibitemOpen
  \bibfield  {author} {\bibinfo {author} {\bibfnamefont {C.}~\bibnamefont
  {Moreau-Luchaire}}, \bibinfo {author} {\bibfnamefont {C.}~\bibnamefont
  {Moutafis}}, \bibinfo {author} {\bibfnamefont {N.}~\bibnamefont {Reyren}},
  \bibinfo {author} {\bibfnamefont {J.}~\bibnamefont {Sampaio}}, \bibinfo
  {author} {\bibfnamefont {C.~A.~F.}\ \bibnamefont {Vaz}}, \bibinfo {author}
  {\bibfnamefont {N.}~\bibnamefont {Van~Horne}}, \bibinfo {author}
  {\bibfnamefont {K.}~\bibnamefont {Bouzehouane}}, \bibinfo {author}
  {\bibfnamefont {K.}~\bibnamefont {Garcia}}, \bibinfo {author} {\bibfnamefont
  {C.}~\bibnamefont {Deranlot}}, \bibinfo {author} {\bibfnamefont
  {P.}~\bibnamefont {Warnicke}}, \bibinfo {author} {\bibfnamefont
  {P.}~\bibnamefont {Wohlh{\"u}ter}}, \bibinfo {author} {\bibfnamefont {J.-M.}\
  \bibnamefont {George}}, \bibinfo {author} {\bibfnamefont {M.}~\bibnamefont
  {Weigand}}, \bibinfo {author} {\bibfnamefont {J.}~\bibnamefont {Raabe}},
  \bibinfo {author} {\bibfnamefont {V.}~\bibnamefont {Cros}},\ and\ \bibinfo
  {author} {\bibfnamefont {A.}~\bibnamefont {Fert}},\ }\href
  {https://doi.org/10.1038/nnano.2015.313} {\bibfield  {journal} {\bibinfo
  {journal} {Nature Nanotechnology}\ }\textbf {\bibinfo {volume} {11}},\
  \bibinfo {pages} {444} (\bibinfo {year} {2016})}\BibitemShut {NoStop}%
\bibitem [{\citenamefont {Yang}\ \emph
  {et~al.}(2018{\natexlab{a}})\citenamefont {Yang}, \citenamefont {Boulle},
  \citenamefont {Cros}, \citenamefont {Fert},\ and\ \citenamefont
  {Chshiev}}]{bib:yang18b}%
  \BibitemOpen
  \bibfield  {author} {\bibinfo {author} {\bibfnamefont {H.}~\bibnamefont
  {Yang}}, \bibinfo {author} {\bibfnamefont {O.}~\bibnamefont {Boulle}},
  \bibinfo {author} {\bibfnamefont {V.}~\bibnamefont {Cros}}, \bibinfo {author}
  {\bibfnamefont {A.}~\bibnamefont {Fert}},\ and\ \bibinfo {author}
  {\bibfnamefont {M.}~\bibnamefont {Chshiev}},\ }\href
  {https://doi.org/10.1038/s41598-018-30063-y} {\bibfield  {journal} {\bibinfo
  {journal} {Scientific Reports}\ }\textbf {\bibinfo {volume} {8}},\ \bibinfo
  {pages} {12356} (\bibinfo {year} {2018}{\natexlab{a}})}\BibitemShut {NoStop}%
\bibitem [{\citenamefont {Perini}\ \emph {et~al.}(2018)\citenamefont {Perini},
  \citenamefont {Meyer}, \citenamefont {Dup\'e}, \citenamefont {von Malottki},
  \citenamefont {Kubetzka}, \citenamefont {von Bergmann}, \citenamefont
  {Wiesendanger},\ and\ \citenamefont {Heinze}}]{bib:perini18}%
  \BibitemOpen
  \bibfield  {author} {\bibinfo {author} {\bibfnamefont {M.}~\bibnamefont
  {Perini}}, \bibinfo {author} {\bibfnamefont {S.}~\bibnamefont {Meyer}},
  \bibinfo {author} {\bibfnamefont {B.}~\bibnamefont {Dup\'e}}, \bibinfo
  {author} {\bibfnamefont {S.}~\bibnamefont {von Malottki}}, \bibinfo {author}
  {\bibfnamefont {A.}~\bibnamefont {Kubetzka}}, \bibinfo {author}
  {\bibfnamefont {K.}~\bibnamefont {von Bergmann}}, \bibinfo {author}
  {\bibfnamefont {R.}~\bibnamefont {Wiesendanger}},\ and\ \bibinfo {author}
  {\bibfnamefont {S.}~\bibnamefont {Heinze}},\ }\href
  {https://doi.org/10.1103/PhysRevB.97.184425} {\bibfield  {journal} {\bibinfo
  {journal} {Phys. Rev. B}\ }\textbf {\bibinfo {volume} {97}},\ \bibinfo
  {pages} {184425} (\bibinfo {year} {2018})}\BibitemShut {NoStop}%
\bibitem [{\citenamefont {Woo}\ \emph {et~al.}(2016)\citenamefont {Woo},
  \citenamefont {Litzius}, \citenamefont {Kr{\"u}ger}, \citenamefont {Im},
  \citenamefont {Caretta}, \citenamefont {Richter}, \citenamefont {Mann},
  \citenamefont {Krone}, \citenamefont {Reeve}, \citenamefont {Weigand},
  \citenamefont {Agrawal}, \citenamefont {Lemesh}, \citenamefont {Mawass},
  \citenamefont {Fischer}, \citenamefont {Kl{\"a}ui},\ and\ \citenamefont
  {Beach}}]{bib:woo16}%
  \BibitemOpen
  \bibfield  {author} {\bibinfo {author} {\bibfnamefont {S.}~\bibnamefont
  {Woo}}, \bibinfo {author} {\bibfnamefont {K.}~\bibnamefont {Litzius}},
  \bibinfo {author} {\bibfnamefont {B.}~\bibnamefont {Kr{\"u}ger}}, \bibinfo
  {author} {\bibfnamefont {M.-Y.}\ \bibnamefont {Im}}, \bibinfo {author}
  {\bibfnamefont {L.}~\bibnamefont {Caretta}}, \bibinfo {author} {\bibfnamefont
  {K.}~\bibnamefont {Richter}}, \bibinfo {author} {\bibfnamefont
  {M.}~\bibnamefont {Mann}}, \bibinfo {author} {\bibfnamefont {A.}~\bibnamefont
  {Krone}}, \bibinfo {author} {\bibfnamefont {R.~M.}\ \bibnamefont {Reeve}},
  \bibinfo {author} {\bibfnamefont {M.}~\bibnamefont {Weigand}}, \bibinfo
  {author} {\bibfnamefont {P.}~\bibnamefont {Agrawal}}, \bibinfo {author}
  {\bibfnamefont {I.}~\bibnamefont {Lemesh}}, \bibinfo {author} {\bibfnamefont
  {M.-A.}\ \bibnamefont {Mawass}}, \bibinfo {author} {\bibfnamefont
  {P.}~\bibnamefont {Fischer}}, \bibinfo {author} {\bibfnamefont
  {M.}~\bibnamefont {Kl{\"a}ui}},\ and\ \bibinfo {author} {\bibfnamefont
  {G.~S.~D.}\ \bibnamefont {Beach}},\ }\href {https://doi.org/10.1038/nmat4593}
  {\bibfield  {journal} {\bibinfo  {journal} {Nature Materials}\ }\textbf
  {\bibinfo {volume} {15}},\ \bibinfo {pages} {501} (\bibinfo {year}
  {2016})}\BibitemShut {NoStop}%
\bibitem [{\citenamefont {Yang}\ \emph {et~al.}(2015)\citenamefont {Yang},
  \citenamefont {Thiaville}, \citenamefont {Rohart}, \citenamefont {Fert},\
  and\ \citenamefont {Chshiev}}]{bib:yang15}%
  \BibitemOpen
  \bibfield  {author} {\bibinfo {author} {\bibfnamefont {H.}~\bibnamefont
  {Yang}}, \bibinfo {author} {\bibfnamefont {A.}~\bibnamefont {Thiaville}},
  \bibinfo {author} {\bibfnamefont {S.}~\bibnamefont {Rohart}}, \bibinfo
  {author} {\bibfnamefont {A.}~\bibnamefont {Fert}},\ and\ \bibinfo {author}
  {\bibfnamefont {M.}~\bibnamefont {Chshiev}},\ }\href
  {https://doi.org/10.1103/PhysRevLett.115.267210} {\bibfield  {journal}
  {\bibinfo  {journal} {Phys. Rev. Lett.}\ }\textbf {\bibinfo {volume} {115}},\
  \bibinfo {pages} {267210} (\bibinfo {year} {2015})}\BibitemShut {NoStop}%
\bibitem [{\citenamefont {Belabbes}\ \emph
  {et~al.}(2016{\natexlab{a}})\citenamefont {Belabbes}, \citenamefont
  {Bihlmayer}, \citenamefont {Bechstedt}, \citenamefont {Bl\"ugel},\ and\
  \citenamefont {Manchon}}]{bib:belabbes16}%
  \BibitemOpen
  \bibfield  {author} {\bibinfo {author} {\bibfnamefont {A.}~\bibnamefont
  {Belabbes}}, \bibinfo {author} {\bibfnamefont {G.}~\bibnamefont {Bihlmayer}},
  \bibinfo {author} {\bibfnamefont {F.}~\bibnamefont {Bechstedt}}, \bibinfo
  {author} {\bibfnamefont {S.}~\bibnamefont {Bl\"ugel}},\ and\ \bibinfo
  {author} {\bibfnamefont {A.}~\bibnamefont {Manchon}},\ }\href
  {https://doi.org/10.1103/PhysRevLett.117.247202} {\bibfield  {journal}
  {\bibinfo  {journal} {Phys. Rev. Lett.}\ }\textbf {\bibinfo {volume} {117}},\
  \bibinfo {pages} {247202} (\bibinfo {year} {2016}{\natexlab{a}})}\BibitemShut
  {NoStop}%
\bibitem [{\citenamefont {Sandratskii}(2017)}]{bib:sandratskii17}%
  \BibitemOpen
  \bibfield  {author} {\bibinfo {author} {\bibfnamefont {L.~M.}\ \bibnamefont
  {Sandratskii}},\ }\href {https://doi.org/10.1103/PhysRevB.96.024450}
  {\bibfield  {journal} {\bibinfo  {journal} {Phys. Rev. B}\ }\textbf {\bibinfo
  {volume} {96}},\ \bibinfo {pages} {024450} (\bibinfo {year}
  {2017})}\BibitemShut {NoStop}%
\bibitem [{\citenamefont {Chen}\ \emph {et~al.}(2021)\citenamefont {Chen},
  \citenamefont {Robertson}, \citenamefont {Hoffmann}, \citenamefont {Ophus},
  \citenamefont {Fernandes~Cauduro}, \citenamefont {Lo~Conte}, \citenamefont
  {Ding}, \citenamefont {Wiesendanger}, \citenamefont {Bl\"ugel}, \citenamefont
  {Schmid},\ and\ \citenamefont {Liu}}]{bib:chen21}%
  \BibitemOpen
  \bibfield  {author} {\bibinfo {author} {\bibfnamefont {G.}~\bibnamefont
  {Chen}}, \bibinfo {author} {\bibfnamefont {M.}~\bibnamefont {Robertson}},
  \bibinfo {author} {\bibfnamefont {M.}~\bibnamefont {Hoffmann}}, \bibinfo
  {author} {\bibfnamefont {C.}~\bibnamefont {Ophus}}, \bibinfo {author}
  {\bibfnamefont {A.~L.}\ \bibnamefont {Fernandes~Cauduro}}, \bibinfo {author}
  {\bibfnamefont {R.}~\bibnamefont {Lo~Conte}}, \bibinfo {author}
  {\bibfnamefont {H.}~\bibnamefont {Ding}}, \bibinfo {author} {\bibfnamefont
  {R.}~\bibnamefont {Wiesendanger}}, \bibinfo {author} {\bibfnamefont
  {S.}~\bibnamefont {Bl\"ugel}}, \bibinfo {author} {\bibfnamefont {A.~K.}\
  \bibnamefont {Schmid}},\ and\ \bibinfo {author} {\bibfnamefont
  {K.}~\bibnamefont {Liu}},\ }\href
  {https://doi.org/10.1103/PhysRevX.11.021015} {\bibfield  {journal} {\bibinfo
  {journal} {Phys. Rev. X}\ }\textbf {\bibinfo {volume} {11}},\ \bibinfo
  {pages} {021015} (\bibinfo {year} {2021})}\BibitemShut {NoStop}%
\bibitem [{\citenamefont {Yang}\ \emph {et~al.}(2020)\citenamefont {Yang},
  \citenamefont {Cui}, \citenamefont {Liang}, \citenamefont {Chshiev},\ and\
  \citenamefont {Yang}}]{bib:yang20}%
  \BibitemOpen
  \bibfield  {author} {\bibinfo {author} {\bibfnamefont {B.}~\bibnamefont
  {Yang}}, \bibinfo {author} {\bibfnamefont {Q.}~\bibnamefont {Cui}}, \bibinfo
  {author} {\bibfnamefont {J.}~\bibnamefont {Liang}}, \bibinfo {author}
  {\bibfnamefont {M.}~\bibnamefont {Chshiev}},\ and\ \bibinfo {author}
  {\bibfnamefont {H.}~\bibnamefont {Yang}},\ }\href
  {https://doi.org/10.1103/PhysRevB.101.014406} {\bibfield  {journal} {\bibinfo
   {journal} {Phys. Rev. B}\ }\textbf {\bibinfo {volume} {101}},\ \bibinfo
  {pages} {014406} (\bibinfo {year} {2020})}\BibitemShut {NoStop}%
\bibitem [{\citenamefont {Belabbes}\ \emph
  {et~al.}(2016{\natexlab{b}})\citenamefont {Belabbes}, \citenamefont
  {Bihlmayer}, \citenamefont {Bl{\"u}gel},\ and\ \citenamefont
  {Manchon}}]{bib:belabbes16b}%
  \BibitemOpen
  \bibfield  {author} {\bibinfo {author} {\bibfnamefont {A.}~\bibnamefont
  {Belabbes}}, \bibinfo {author} {\bibfnamefont {G.}~\bibnamefont {Bihlmayer}},
  \bibinfo {author} {\bibfnamefont {S.}~\bibnamefont {Bl{\"u}gel}},\ and\
  \bibinfo {author} {\bibfnamefont {A.}~\bibnamefont {Manchon}},\ }\href
  {https://doi.org/10.1038/srep24634} {\bibfield  {journal} {\bibinfo
  {journal} {Scientific Reports}\ }\textbf {\bibinfo {volume} {6}},\ \bibinfo
  {pages} {24634} (\bibinfo {year} {2016}{\natexlab{b}})}\BibitemShut {NoStop}%
\bibitem [{\citenamefont {Chen}\ \emph {et~al.}(2020)\citenamefont {Chen},
  \citenamefont {Mascaraque}, \citenamefont {Jia}, \citenamefont {Zimmermann},
  \citenamefont {Robertson}, \citenamefont {Conte}, \citenamefont {Hoffmann},
  \citenamefont {Gonz{\'a}lez~Barrio}, \citenamefont {Ding}, \citenamefont
  {Wiesendanger}, \citenamefont {Michel}, \citenamefont {Bl{\"u}gel},
  \citenamefont {Schmid},\ and\ \citenamefont {Liu}}]{bib:chen20}%
  \BibitemOpen
  \bibfield  {author} {\bibinfo {author} {\bibfnamefont {G.}~\bibnamefont
  {Chen}}, \bibinfo {author} {\bibfnamefont {A.}~\bibnamefont {Mascaraque}},
  \bibinfo {author} {\bibfnamefont {H.}~\bibnamefont {Jia}}, \bibinfo {author}
  {\bibfnamefont {B.}~\bibnamefont {Zimmermann}}, \bibinfo {author}
  {\bibfnamefont {M.}~\bibnamefont {Robertson}}, \bibinfo {author}
  {\bibfnamefont {R.~L.}\ \bibnamefont {Conte}}, \bibinfo {author}
  {\bibfnamefont {M.}~\bibnamefont {Hoffmann}}, \bibinfo {author}
  {\bibfnamefont {M.~A.}\ \bibnamefont {Gonz{\'a}lez~Barrio}}, \bibinfo
  {author} {\bibfnamefont {H.}~\bibnamefont {Ding}}, \bibinfo {author}
  {\bibfnamefont {R.}~\bibnamefont {Wiesendanger}}, \bibinfo {author}
  {\bibfnamefont {E.~G.}\ \bibnamefont {Michel}}, \bibinfo {author}
  {\bibfnamefont {S.}~\bibnamefont {Bl{\"u}gel}}, \bibinfo {author}
  {\bibfnamefont {A.~K.}\ \bibnamefont {Schmid}},\ and\ \bibinfo {author}
  {\bibfnamefont {K.}~\bibnamefont {Liu}},\ }\bibfield  {journal} {\bibinfo
  {journal} {Science Advances}\ }\textbf {\bibinfo {volume} {6}},\ \href
  {https://doi.org/10.1126/sciadv.aba4924} {10.1126/sciadv.aba4924} (\bibinfo
  {year} {2020}),\ \Eprint
  {https://arxiv.org/abs/https://advances.sciencemag.org/content/6/33/eaba4924.full.pdf}
  {https://advances.sciencemag.org/content/6/33/eaba4924.full.pdf} \BibitemShut
  {NoStop}%
\bibitem [{\citenamefont {Ajejas}\ \emph {et~al.}(2018)\citenamefont {Ajejas},
  \citenamefont {Gud{\'\i}n}, \citenamefont {Guerrero}, \citenamefont
  {Anad{\'o}n~Barcelona}, \citenamefont {Diez}, \citenamefont {de~Melo~Costa},
  \citenamefont {Olleros}, \citenamefont {Ni{\~n}o}, \citenamefont {Pizzini},
  \citenamefont {Vogel}, \citenamefont {Valvidares}, \citenamefont {Gargiani},
  \citenamefont {Cabero}, \citenamefont {Varela}, \citenamefont {Camarero},
  \citenamefont {Miranda},\ and\ \citenamefont {Perna}}]{bib:ajejas18}%
  \BibitemOpen
  \bibfield  {author} {\bibinfo {author} {\bibfnamefont {F.}~\bibnamefont
  {Ajejas}}, \bibinfo {author} {\bibfnamefont {A.}~\bibnamefont {Gud{\'\i}n}},
  \bibinfo {author} {\bibfnamefont {R.}~\bibnamefont {Guerrero}}, \bibinfo
  {author} {\bibfnamefont {A.}~\bibnamefont {Anad{\'o}n~Barcelona}}, \bibinfo
  {author} {\bibfnamefont {J.~M.}\ \bibnamefont {Diez}}, \bibinfo {author}
  {\bibfnamefont {L.}~\bibnamefont {de~Melo~Costa}}, \bibinfo {author}
  {\bibfnamefont {P.}~\bibnamefont {Olleros}}, \bibinfo {author} {\bibfnamefont
  {M.~A.}\ \bibnamefont {Ni{\~n}o}}, \bibinfo {author} {\bibfnamefont
  {S.}~\bibnamefont {Pizzini}}, \bibinfo {author} {\bibfnamefont
  {J.}~\bibnamefont {Vogel}}, \bibinfo {author} {\bibfnamefont
  {M.}~\bibnamefont {Valvidares}}, \bibinfo {author} {\bibfnamefont
  {P.}~\bibnamefont {Gargiani}}, \bibinfo {author} {\bibfnamefont
  {M.}~\bibnamefont {Cabero}}, \bibinfo {author} {\bibfnamefont
  {M.}~\bibnamefont {Varela}}, \bibinfo {author} {\bibfnamefont
  {J.}~\bibnamefont {Camarero}}, \bibinfo {author} {\bibfnamefont
  {R.}~\bibnamefont {Miranda}},\ and\ \bibinfo {author} {\bibfnamefont
  {P.}~\bibnamefont {Perna}},\ }\bibfield  {booktitle} {\emph {\bibinfo
  {booktitle} {Nano Letters}},\ }\href
  {https://doi.org/10.1021/acs.nanolett.8b00878} {\bibfield  {journal}
  {\bibinfo  {journal} {Nano Letters}\ }\textbf {\bibinfo {volume} {18}},\
  \bibinfo {pages} {5364} (\bibinfo {year} {2018})}\BibitemShut {NoStop}%
\bibitem [{\citenamefont {Hallal}\ \emph {et~al.}(2021)\citenamefont {Hallal},
  \citenamefont {Liang}, \citenamefont {Ibrahim}, \citenamefont {Yang},
  \citenamefont {Fert},\ and\ \citenamefont {Chshiev}}]{bib:hallal21}%
  \BibitemOpen
  \bibfield  {author} {\bibinfo {author} {\bibfnamefont {A.}~\bibnamefont
  {Hallal}}, \bibinfo {author} {\bibfnamefont {J.}~\bibnamefont {Liang}},
  \bibinfo {author} {\bibfnamefont {F.}~\bibnamefont {Ibrahim}}, \bibinfo
  {author} {\bibfnamefont {H.}~\bibnamefont {Yang}}, \bibinfo {author}
  {\bibfnamefont {A.}~\bibnamefont {Fert}},\ and\ \bibinfo {author}
  {\bibfnamefont {M.}~\bibnamefont {Chshiev}},\ }\bibfield  {journal} {\bibinfo
   {journal} {Nano Letters}\ }\href
  {https://doi.org/10.1021/acs.nanolett.1c01713} {10.1021/acs.nanolett.1c01713}
  (\bibinfo {year} {2021})\BibitemShut {NoStop}%
\bibitem [{\citenamefont {Jia}\ \emph {et~al.}(2020)\citenamefont {Jia},
  \citenamefont {Zimmermann}, \citenamefont {Michalicek}, \citenamefont
  {Bihlmayer},\ and\ \citenamefont {Bl\"ugel}}]{bib:jia20}%
  \BibitemOpen
  \bibfield  {author} {\bibinfo {author} {\bibfnamefont {H.}~\bibnamefont
  {Jia}}, \bibinfo {author} {\bibfnamefont {B.}~\bibnamefont {Zimmermann}},
  \bibinfo {author} {\bibfnamefont {G.}~\bibnamefont {Michalicek}}, \bibinfo
  {author} {\bibfnamefont {G.}~\bibnamefont {Bihlmayer}},\ and\ \bibinfo
  {author} {\bibfnamefont {S.}~\bibnamefont {Bl\"ugel}},\ }\href
  {https://doi.org/10.1103/PhysRevMaterials.4.024405} {\bibfield  {journal}
  {\bibinfo  {journal} {Phys. Rev. Materials}\ }\textbf {\bibinfo {volume}
  {4}},\ \bibinfo {pages} {024405} (\bibinfo {year} {2020})}\BibitemShut
  {NoStop}%
\bibitem [{\citenamefont {Kim}\ \emph {et~al.}(2018{\natexlab{a}})\citenamefont
  {Kim}, \citenamefont {Ueda}, \citenamefont {Go}, \citenamefont {Jang},
  \citenamefont {Lee}, \citenamefont {Belabbes}, \citenamefont {Manchon},
  \citenamefont {Suzuki}, \citenamefont {Kotani}, \citenamefont {Nakamura},
  \citenamefont {Nakamura}, \citenamefont {Koyama}, \citenamefont {Chiba},
  \citenamefont {Yamada}, \citenamefont {Kim}, \citenamefont {Moriyama},
  \citenamefont {Kim},\ and\ \citenamefont {Ono}}]{bib:kim18b}%
  \BibitemOpen
  \bibfield  {author} {\bibinfo {author} {\bibfnamefont {S.}~\bibnamefont
  {Kim}}, \bibinfo {author} {\bibfnamefont {K.}~\bibnamefont {Ueda}}, \bibinfo
  {author} {\bibfnamefont {G.}~\bibnamefont {Go}}, \bibinfo {author}
  {\bibfnamefont {P.-H.}\ \bibnamefont {Jang}}, \bibinfo {author}
  {\bibfnamefont {K.-J.}\ \bibnamefont {Lee}}, \bibinfo {author} {\bibfnamefont
  {A.}~\bibnamefont {Belabbes}}, \bibinfo {author} {\bibfnamefont
  {A.}~\bibnamefont {Manchon}}, \bibinfo {author} {\bibfnamefont
  {M.}~\bibnamefont {Suzuki}}, \bibinfo {author} {\bibfnamefont
  {Y.}~\bibnamefont {Kotani}}, \bibinfo {author} {\bibfnamefont
  {T.}~\bibnamefont {Nakamura}}, \bibinfo {author} {\bibfnamefont
  {K.}~\bibnamefont {Nakamura}}, \bibinfo {author} {\bibfnamefont
  {T.}~\bibnamefont {Koyama}}, \bibinfo {author} {\bibfnamefont
  {D.}~\bibnamefont {Chiba}}, \bibinfo {author} {\bibfnamefont {K.~T.}\
  \bibnamefont {Yamada}}, \bibinfo {author} {\bibfnamefont {D.-H.}\
  \bibnamefont {Kim}}, \bibinfo {author} {\bibfnamefont {T.}~\bibnamefont
  {Moriyama}}, \bibinfo {author} {\bibfnamefont {K.-J.}\ \bibnamefont {Kim}},\
  and\ \bibinfo {author} {\bibfnamefont {T.}~\bibnamefont {Ono}},\ }\href
  {https://doi.org/10.1038/s41467-018-04017-x} {\bibfield  {journal} {\bibinfo
  {journal} {Nature Communications}\ }\textbf {\bibinfo {volume} {9}},\
  \bibinfo {pages} {1648} (\bibinfo {year} {2018}{\natexlab{a}})}\BibitemShut
  {NoStop}%
\bibitem [{\citenamefont {Hsu}\ \emph {et~al.}(2017)\citenamefont {Hsu},
  \citenamefont {Kubetzka}, \citenamefont {Finco}, \citenamefont {Romming},
  \citenamefont {von Bergmann},\ and\ \citenamefont
  {Wiesendanger}}]{bib:hsu17}%
  \BibitemOpen
  \bibfield  {author} {\bibinfo {author} {\bibfnamefont {P.-J.}\ \bibnamefont
  {Hsu}}, \bibinfo {author} {\bibfnamefont {A.}~\bibnamefont {Kubetzka}},
  \bibinfo {author} {\bibfnamefont {A.}~\bibnamefont {Finco}}, \bibinfo
  {author} {\bibfnamefont {N.}~\bibnamefont {Romming}}, \bibinfo {author}
  {\bibfnamefont {K.}~\bibnamefont {von Bergmann}},\ and\ \bibinfo {author}
  {\bibfnamefont {R.}~\bibnamefont {Wiesendanger}},\ }\href
  {https://doi.org/10.1038/nnano.2016.234} {\bibfield  {journal} {\bibinfo
  {journal} {Nature Nanotechnology}\ }\textbf {\bibinfo {volume} {12}},\
  \bibinfo {pages} {123} (\bibinfo {year} {2017})}\BibitemShut {NoStop}%
\bibitem [{\citenamefont {Yang}\ \emph
  {et~al.}(2018{\natexlab{b}})\citenamefont {Yang}, \citenamefont {Chen},
  \citenamefont {Cotta}, \citenamefont {N'Diaye}, \citenamefont {Nikolaev},
  \citenamefont {Soares}, \citenamefont {Macedo}, \citenamefont {Liu},
  \citenamefont {Schmid}, \citenamefont {Fert},\ and\ \citenamefont
  {Chshiev}}]{bib:yang18}%
  \BibitemOpen
  \bibfield  {author} {\bibinfo {author} {\bibfnamefont {H.}~\bibnamefont
  {Yang}}, \bibinfo {author} {\bibfnamefont {G.}~\bibnamefont {Chen}}, \bibinfo
  {author} {\bibfnamefont {A.~A.~C.}\ \bibnamefont {Cotta}}, \bibinfo {author}
  {\bibfnamefont {A.~T.}\ \bibnamefont {N'Diaye}}, \bibinfo {author}
  {\bibfnamefont {S.~A.}\ \bibnamefont {Nikolaev}}, \bibinfo {author}
  {\bibfnamefont {E.~A.}\ \bibnamefont {Soares}}, \bibinfo {author}
  {\bibfnamefont {W.~A.~A.}\ \bibnamefont {Macedo}}, \bibinfo {author}
  {\bibfnamefont {K.}~\bibnamefont {Liu}}, \bibinfo {author} {\bibfnamefont
  {A.~K.}\ \bibnamefont {Schmid}}, \bibinfo {author} {\bibfnamefont
  {A.}~\bibnamefont {Fert}},\ and\ \bibinfo {author} {\bibfnamefont
  {M.}~\bibnamefont {Chshiev}},\ }\href
  {https://doi.org/10.1038/s41563-018-0079-4} {\bibfield  {journal} {\bibinfo
  {journal} {Nature Materials}\ }\textbf {\bibinfo {volume} {17}},\ \bibinfo
  {pages} {605} (\bibinfo {year} {2018}{\natexlab{b}})}\BibitemShut {NoStop}%
\bibitem [{\citenamefont {Imamura}\ \emph {et~al.}(2004)\citenamefont
  {Imamura}, \citenamefont {Bruno},\ and\ \citenamefont
  {Utsumi}}]{bib:imamura04}%
  \BibitemOpen
  \bibfield  {author} {\bibinfo {author} {\bibfnamefont {H.}~\bibnamefont
  {Imamura}}, \bibinfo {author} {\bibfnamefont {P.}~\bibnamefont {Bruno}},\
  and\ \bibinfo {author} {\bibfnamefont {Y.}~\bibnamefont {Utsumi}},\ }\href
  {https://doi.org/10.1103/PhysRevB.69.121303} {\bibfield  {journal} {\bibinfo
  {journal} {Phys. Rev. B}\ }\textbf {\bibinfo {volume} {69}},\ \bibinfo
  {pages} {121303} (\bibinfo {year} {2004})}\BibitemShut {NoStop}%
\bibitem [{\citenamefont {Kim}\ \emph {et~al.}(2013)\citenamefont {Kim},
  \citenamefont {Lee}, \citenamefont {Lee},\ and\ \citenamefont
  {Stiles}}]{bib:kim13}%
  \BibitemOpen
  \bibfield  {author} {\bibinfo {author} {\bibfnamefont {K.-W.}\ \bibnamefont
  {Kim}}, \bibinfo {author} {\bibfnamefont {H.-W.}\ \bibnamefont {Lee}},
  \bibinfo {author} {\bibfnamefont {K.-J.}\ \bibnamefont {Lee}},\ and\ \bibinfo
  {author} {\bibfnamefont {M.~D.}\ \bibnamefont {Stiles}},\ }\href
  {https://doi.org/10.1103/PhysRevLett.111.216601} {\bibfield  {journal}
  {\bibinfo  {journal} {Phys. Rev. Lett.}\ }\textbf {\bibinfo {volume} {111}},\
  \bibinfo {pages} {216601} (\bibinfo {year} {2013})}\BibitemShut {NoStop}%
\bibitem [{\citenamefont {Kundu}\ and\ \citenamefont
  {Zhang}(2015)}]{bib:kundu15}%
  \BibitemOpen
  \bibfield  {author} {\bibinfo {author} {\bibfnamefont {A.}~\bibnamefont
  {Kundu}}\ and\ \bibinfo {author} {\bibfnamefont {S.}~\bibnamefont {Zhang}},\
  }\href {https://doi.org/10.1103/PhysRevB.92.094434} {\bibfield  {journal}
  {\bibinfo  {journal} {Phys. Rev. B}\ }\textbf {\bibinfo {volume} {92}},\
  \bibinfo {pages} {094434} (\bibinfo {year} {2015})}\BibitemShut {NoStop}%
\bibitem [{\citenamefont {Blanco-Rey}\ \emph {et~al.}(2021)\citenamefont
  {Blanco-Rey}, \citenamefont {Perna}, \citenamefont {Gudin}, \citenamefont
  {Diez}, \citenamefont {Anadon}, \citenamefont {Olleros-Rodriguez},
  \citenamefont {de~Melo~Costa}, \citenamefont {Valvidares}, \citenamefont
  {Gargiani}, \citenamefont {Guedeja-Marron}, \citenamefont {Cabero},
  \citenamefont {Varela}, \citenamefont {Garcia-Fernandez}, \citenamefont
  {Otrokov}, \citenamefont {Camarero}, \citenamefont {Miranda}, \citenamefont
  {Arnau},\ and\ \citenamefont {Cerda}}]{bib:blanco21}%
  \BibitemOpen
  \bibfield  {author} {\bibinfo {author} {\bibfnamefont {M.}~\bibnamefont
  {Blanco-Rey}}, \bibinfo {author} {\bibfnamefont {P.}~\bibnamefont {Perna}},
  \bibinfo {author} {\bibfnamefont {A.}~\bibnamefont {Gudin}}, \bibinfo
  {author} {\bibfnamefont {J.~M.}\ \bibnamefont {Diez}}, \bibinfo {author}
  {\bibfnamefont {A.}~\bibnamefont {Anadon}}, \bibinfo {author} {\bibfnamefont
  {P.}~\bibnamefont {Olleros-Rodriguez}}, \bibinfo {author} {\bibfnamefont
  {L.}~\bibnamefont {de~Melo~Costa}}, \bibinfo {author} {\bibfnamefont
  {M.}~\bibnamefont {Valvidares}}, \bibinfo {author} {\bibfnamefont
  {P.}~\bibnamefont {Gargiani}}, \bibinfo {author} {\bibfnamefont
  {A.}~\bibnamefont {Guedeja-Marron}}, \bibinfo {author} {\bibfnamefont
  {M.}~\bibnamefont {Cabero}}, \bibinfo {author} {\bibfnamefont
  {M.}~\bibnamefont {Varela}}, \bibinfo {author} {\bibfnamefont
  {C.}~\bibnamefont {Garcia-Fernandez}}, \bibinfo {author} {\bibfnamefont
  {M.~M.}\ \bibnamefont {Otrokov}}, \bibinfo {author} {\bibfnamefont
  {J.}~\bibnamefont {Camarero}}, \bibinfo {author} {\bibfnamefont
  {R.}~\bibnamefont {Miranda}}, \bibinfo {author} {\bibfnamefont
  {A.}~\bibnamefont {Arnau}},\ and\ \bibinfo {author} {\bibfnamefont {J.~I.}\
  \bibnamefont {Cerda}},\ }\href {https://doi.org/10.1021/acsanm.0c03364}
  {\bibfield  {journal} {\bibinfo  {journal} {ACS Applied Nano Materials}\
  }\textbf {\bibinfo {volume} {4}},\ \bibinfo {pages} {4398} (\bibinfo {year}
  {2021})}\BibitemShut {NoStop}%
\bibitem [{\citenamefont {Cr\'epieux}\ and\ \citenamefont
  {Lacroix}(1998)}]{bib:crepieux98}%
  \BibitemOpen
  \bibfield  {author} {\bibinfo {author} {\bibfnamefont {A.}~\bibnamefont
  {Cr\'epieux}}\ and\ \bibinfo {author} {\bibfnamefont {C.}~\bibnamefont
  {Lacroix}},\ }\href
  {https://doi.org/https://doi.org/10.1016/S0304-8853(97)01044-5} {\bibfield
  {journal} {\bibinfo  {journal} {Journal of Magnetism and Magnetic Materials}\
  }\textbf {\bibinfo {volume} {182}},\ \bibinfo {pages} {341} (\bibinfo {year}
  {1998})}\BibitemShut {NoStop}%
\bibitem [{\citenamefont {Krakauer}\ \emph {et~al.}(1979)\citenamefont
  {Krakauer}, \citenamefont {Posternak},\ and\ \citenamefont
  {Freeman}}]{bib:krakauer79}%
  \BibitemOpen
  \bibfield  {author} {\bibinfo {author} {\bibfnamefont {H.}~\bibnamefont
  {Krakauer}}, \bibinfo {author} {\bibfnamefont {M.}~\bibnamefont
  {Posternak}},\ and\ \bibinfo {author} {\bibfnamefont {A.~J.}\ \bibnamefont
  {Freeman}},\ }\href {https://doi.org/10.1103/PhysRevB.19.1706} {\bibfield
  {journal} {\bibinfo  {journal} {Phys. Rev. B}\ }\textbf {\bibinfo {volume}
  {19}},\ \bibinfo {pages} {1706} (\bibinfo {year} {1979})}\BibitemShut
  {NoStop}%
\bibitem [{\citenamefont {Wimmer}\ \emph {et~al.}(1981)\citenamefont {Wimmer},
  \citenamefont {Krakauer}, \citenamefont {Weinert},\ and\ \citenamefont
  {Freeman}}]{bib:wimmer81}%
  \BibitemOpen
  \bibfield  {author} {\bibinfo {author} {\bibfnamefont {E.}~\bibnamefont
  {Wimmer}}, \bibinfo {author} {\bibfnamefont {H.}~\bibnamefont {Krakauer}},
  \bibinfo {author} {\bibfnamefont {M.}~\bibnamefont {Weinert}},\ and\ \bibinfo
  {author} {\bibfnamefont {A.~J.}\ \bibnamefont {Freeman}},\ }\href@noop {}
  {\bibfield  {journal} {\bibinfo  {journal} {Phys. Rev. B}\ }\textbf {\bibinfo
  {volume} {24}},\ \bibinfo {pages} {864} (\bibinfo {year} {1981})}\BibitemShut
  {NoStop}%
\bibitem [{bib()}]{bib:fleur}%
  \BibitemOpen
  \href@noop {} {}\bibinfo {note} {\textsc{Fleur} site:
  http://www.flapw.de}\BibitemShut {NoStop}%
\bibitem [{\citenamefont {Perdew}\ \emph {et~al.}(1996)\citenamefont {Perdew},
  \citenamefont {Burke},\ and\ \citenamefont {Ernzerhof}}]{bib:pbe96}%
  \BibitemOpen
  \bibfield  {author} {\bibinfo {author} {\bibfnamefont {J.~P.}\ \bibnamefont
  {Perdew}}, \bibinfo {author} {\bibfnamefont {K.}~\bibnamefont {Burke}},\ and\
  \bibinfo {author} {\bibfnamefont {M.}~\bibnamefont {Ernzerhof}},\ }\href
  {https://doi.org/10.1103/PhysRevLett.77.3865} {\bibfield  {journal} {\bibinfo
   {journal} {Phys. Rev. Lett.}\ }\textbf {\bibinfo {volume} {77}},\ \bibinfo
  {pages} {3865} (\bibinfo {year} {1996})}\BibitemShut {NoStop}%
\bibitem [{\citenamefont {Monkhorst}\ and\ \citenamefont
  {Pack}(1976)}]{bib:monk76}%
  \BibitemOpen
  \bibfield  {author} {\bibinfo {author} {\bibfnamefont {H.~J.}\ \bibnamefont
  {Monkhorst}}\ and\ \bibinfo {author} {\bibfnamefont {J.~D.}\ \bibnamefont
  {Pack}},\ }\href {https://doi.org/10.1103/PhysRevB.13.5188} {\bibfield
  {journal} {\bibinfo  {journal} {Phys. Rev. B}\ }\textbf {\bibinfo {volume}
  {13}},\ \bibinfo {pages} {5188} (\bibinfo {year} {1976})}\BibitemShut
  {NoStop}%
\bibitem [{\citenamefont {Sandratskii}(1991)}]{bib:sandratskii91}%
  \BibitemOpen
  \bibfield  {author} {\bibinfo {author} {\bibfnamefont {L.~M.}\ \bibnamefont
  {Sandratskii}},\ }\href {https://doi.org/10.1088/0953-8984/3/44/004}
  {\bibfield  {journal} {\bibinfo  {journal} {Journal of Physics: Condensed
  Matter}\ }\textbf {\bibinfo {volume} {3}},\ \bibinfo {pages} {8565} (\bibinfo
  {year} {1991})}\BibitemShut {NoStop}%
\bibitem [{\citenamefont {Heide}\ \emph {et~al.}(2009)\citenamefont {Heide},
  \citenamefont {Bihlmayer},\ and\ \citenamefont {Blügel}}]{bib:heide09}%
  \BibitemOpen
  \bibfield  {author} {\bibinfo {author} {\bibfnamefont {M.}~\bibnamefont
  {Heide}}, \bibinfo {author} {\bibfnamefont {G.}~\bibnamefont {Bihlmayer}},\
  and\ \bibinfo {author} {\bibfnamefont {S.}~\bibnamefont {Blügel}},\ }\href
  {https://doi.org/https://doi.org/10.1016/j.physb.2009.06.070} {\bibfield
  {journal} {\bibinfo  {journal} {Physica B: Condensed Matter}\ }\textbf
  {\bibinfo {volume} {404}},\ \bibinfo {pages} {2678} (\bibinfo {year}
  {2009})}\BibitemShut {NoStop}%
\bibitem [{\citenamefont {Zimmermann}\ \emph {et~al.}(2019)\citenamefont
  {Zimmermann}, \citenamefont {Bihlmayer}, \citenamefont {B\"ottcher},
  \citenamefont {Bouhassoune}, \citenamefont {Lounis}, \citenamefont {Sinova},
  \citenamefont {Heinze}, \citenamefont {Bl\"ugel},\ and\ \citenamefont
  {Dup\'e}}]{bib:zimmermann19}%
  \BibitemOpen
  \bibfield  {author} {\bibinfo {author} {\bibfnamefont {B.}~\bibnamefont
  {Zimmermann}}, \bibinfo {author} {\bibfnamefont {G.}~\bibnamefont
  {Bihlmayer}}, \bibinfo {author} {\bibfnamefont {M.}~\bibnamefont
  {B\"ottcher}}, \bibinfo {author} {\bibfnamefont {M.}~\bibnamefont
  {Bouhassoune}}, \bibinfo {author} {\bibfnamefont {S.}~\bibnamefont {Lounis}},
  \bibinfo {author} {\bibfnamefont {J.}~\bibnamefont {Sinova}}, \bibinfo
  {author} {\bibfnamefont {S.}~\bibnamefont {Heinze}}, \bibinfo {author}
  {\bibfnamefont {S.}~\bibnamefont {Bl\"ugel}},\ and\ \bibinfo {author}
  {\bibfnamefont {B.}~\bibnamefont {Dup\'e}},\ }\href
  {https://doi.org/10.1103/PhysRevB.99.214426} {\bibfield  {journal} {\bibinfo
  {journal} {Phys. Rev. B}\ }\textbf {\bibinfo {volume} {99}},\ \bibinfo
  {pages} {214426} (\bibinfo {year} {2019})}\BibitemShut {NoStop}%
\bibitem [{\citenamefont {Lee}\ \emph {et~al.}(2009)\citenamefont {Lee},
  \citenamefont {Behera}, \citenamefont {Wu}, \citenamefont {Xu}, \citenamefont
  {Li}, \citenamefont {Sinnott}, \citenamefont {Phillpot}, \citenamefont
  {Chen},\ and\ \citenamefont {Gopalan}}]{bib:lee09}%
  \BibitemOpen
  \bibfield  {author} {\bibinfo {author} {\bibfnamefont {D.}~\bibnamefont
  {Lee}}, \bibinfo {author} {\bibfnamefont {R.~K.}\ \bibnamefont {Behera}},
  \bibinfo {author} {\bibfnamefont {P.}~\bibnamefont {Wu}}, \bibinfo {author}
  {\bibfnamefont {H.}~\bibnamefont {Xu}}, \bibinfo {author} {\bibfnamefont
  {Y.~L.}\ \bibnamefont {Li}}, \bibinfo {author} {\bibfnamefont {S.~B.}\
  \bibnamefont {Sinnott}}, \bibinfo {author} {\bibfnamefont {S.~R.}\
  \bibnamefont {Phillpot}}, \bibinfo {author} {\bibfnamefont {L.~Q.}\
  \bibnamefont {Chen}},\ and\ \bibinfo {author} {\bibfnamefont
  {V.}~\bibnamefont {Gopalan}},\ }\href
  {https://doi.org/10.1103/PhysRevB.80.060102} {\bibfield  {journal} {\bibinfo
  {journal} {Phys. Rev. B}\ }\textbf {\bibinfo {volume} {80}},\ \bibinfo
  {pages} {060102} (\bibinfo {year} {2009})}\BibitemShut {NoStop}%
\bibitem [{\citenamefont {Kim}\ \emph {et~al.}(2018{\natexlab{b}})\citenamefont
  {Kim}, \citenamefont {Moon}, \citenamefont {Kerber}, \citenamefont
  {Nothhelfer},\ and\ \citenamefont {Everschor-Sitte}}]{bib:kim18}%
  \BibitemOpen
  \bibfield  {author} {\bibinfo {author} {\bibfnamefont {K.-W.}\ \bibnamefont
  {Kim}}, \bibinfo {author} {\bibfnamefont {K.-W.}\ \bibnamefont {Moon}},
  \bibinfo {author} {\bibfnamefont {N.}~\bibnamefont {Kerber}}, \bibinfo
  {author} {\bibfnamefont {J.}~\bibnamefont {Nothhelfer}},\ and\ \bibinfo
  {author} {\bibfnamefont {K.}~\bibnamefont {Everschor-Sitte}},\ }\href
  {https://doi.org/10.1103/PhysRevB.97.224427} {\bibfield  {journal} {\bibinfo
  {journal} {Phys. Rev. B}\ }\textbf {\bibinfo {volume} {97}},\ \bibinfo
  {pages} {224427} (\bibinfo {year} {2018}{\natexlab{b}})}\BibitemShut
  {NoStop}%
\bibitem [{\citenamefont {Blanco-Rey}\ \emph {et~al.}(2019)\citenamefont
  {Blanco-Rey}, \citenamefont {Cerd{\'{a}}},\ and\ \citenamefont
  {Arnau}}]{bib:blanco19}%
  \BibitemOpen
  \bibfield  {author} {\bibinfo {author} {\bibfnamefont {M.}~\bibnamefont
  {Blanco-Rey}}, \bibinfo {author} {\bibfnamefont {J.~I.}\ \bibnamefont
  {Cerd{\'{a}}}},\ and\ \bibinfo {author} {\bibfnamefont {A.}~\bibnamefont
  {Arnau}},\ }\href {https://doi.org/10.1088/1367-2630/ab3060} {\bibfield
  {journal} {\bibinfo  {journal} {New Journal of Physics}\ }\textbf {\bibinfo
  {volume} {21}},\ \bibinfo {pages} {073054} (\bibinfo {year}
  {2019})}\BibitemShut {NoStop}%
\bibitem [{\citenamefont {Ajejas}\ \emph {et~al.}(2020)\citenamefont {Ajejas},
  \citenamefont {Anadon}, \citenamefont {Gudin}, \citenamefont {Diez},
  \citenamefont {Ayani}, \citenamefont {Olleros-Rodr\'{\i}guez}, \citenamefont
  {de~Melo~Costa}, \citenamefont {Nav\'{\i}o}, \citenamefont {Guti\'errez},
  \citenamefont {Calleja}, \citenamefont {V\'azquez~de Parga}, \citenamefont
  {Miranda}, \citenamefont {Camarero},\ and\ \citenamefont
  {Perna}}]{bib:ajejas20}%
  \BibitemOpen
  \bibfield  {author} {\bibinfo {author} {\bibfnamefont {F.}~\bibnamefont
  {Ajejas}}, \bibinfo {author} {\bibfnamefont {A.}~\bibnamefont {Anadon}},
  \bibinfo {author} {\bibfnamefont {A.}~\bibnamefont {Gudin}}, \bibinfo
  {author} {\bibfnamefont {J.~M.}\ \bibnamefont {Diez}}, \bibinfo {author}
  {\bibfnamefont {C.~G.}\ \bibnamefont {Ayani}}, \bibinfo {author}
  {\bibfnamefont {P.}~\bibnamefont {Olleros-Rodr\'{\i}guez}}, \bibinfo {author}
  {\bibfnamefont {L.}~\bibnamefont {de~Melo~Costa}}, \bibinfo {author}
  {\bibfnamefont {C.}~\bibnamefont {Nav\'{\i}o}}, \bibinfo {author}
  {\bibfnamefont {A.}~\bibnamefont {Guti\'errez}}, \bibinfo {author}
  {\bibfnamefont {F.}~\bibnamefont {Calleja}}, \bibinfo {author} {\bibfnamefont
  {A.~L.}\ \bibnamefont {V\'azquez~de Parga}}, \bibinfo {author} {\bibfnamefont
  {R.}~\bibnamefont {Miranda}}, \bibinfo {author} {\bibfnamefont
  {J.}~\bibnamefont {Camarero}},\ and\ \bibinfo {author} {\bibfnamefont
  {P.}~\bibnamefont {Perna}},\ }\href {https://doi.org/10.1021/acsami.9b19159}
  {\bibfield  {journal} {\bibinfo  {journal} {ACS Applied Materials \&
  Interfaces}\ }\textbf {\bibinfo {volume} {12}},\ \bibinfo {pages} {4088}
  (\bibinfo {year} {2020})}\BibitemShut {NoStop}%
\bibitem [{\citenamefont {Daalderop}\ \emph {et~al.}(1994)\citenamefont
  {Daalderop}, \citenamefont {Kelly},\ and\ \citenamefont
  {Schuurmans}}]{bib:daalderop94}%
  \BibitemOpen
  \bibfield  {author} {\bibinfo {author} {\bibfnamefont {G.~H.~O.}\
  \bibnamefont {Daalderop}}, \bibinfo {author} {\bibfnamefont {P.~J.}\
  \bibnamefont {Kelly}},\ and\ \bibinfo {author} {\bibfnamefont {M.~F.~H.}\
  \bibnamefont {Schuurmans}},\ }\href
  {https://doi.org/10.1103/PhysRevB.50.9989} {\bibfield  {journal} {\bibinfo
  {journal} {Phys. Rev. B}\ }\textbf {\bibinfo {volume} {50}},\ \bibinfo
  {pages} {9989} (\bibinfo {year} {1994})}\BibitemShut {NoStop}%
\bibitem [{Note1()}]{Note1}%
  \BibitemOpen
  \bibinfo {note} {Some selection rules are found concerning the orbital
  character $|l,m \rangle $ of the states with wavevectors $\protect \mathbf
  {k}$ and $\protect \mathbf {k} \pm \protect \mathbf {q}$, When the $\protect
  \mathbf {q}$-dependence is perturbatively considered in the analytical
  $D$-vector expression, a dipolar term appears that couples different
  $l$-channels, as shown in Ref.~\cite {bib:jia20}. If only the inversion
  symmetry breaking is considered, as in Ref.~\cite {bib:yang20}, only the
  couplings between $|2,m\rangle $ orbitals of suitable symmetry (determined by
  the SOC term in the hamiltonian \cite {bib:abate65}) survive.}\BibitemShut
  {Stop}%
\bibitem [{\citenamefont {Krupin}\ \emph {et~al.}(2005)\citenamefont {Krupin},
  \citenamefont {Bihlmayer}, \citenamefont {Starke}, \citenamefont {Gorovikov},
  \citenamefont {Prieto}, \citenamefont {D\"obrich}, \citenamefont {Bl\"ugel},\
  and\ \citenamefont {Kaindl}}]{bib:krupin05}%
  \BibitemOpen
  \bibfield  {author} {\bibinfo {author} {\bibfnamefont {O.}~\bibnamefont
  {Krupin}}, \bibinfo {author} {\bibfnamefont {G.}~\bibnamefont {Bihlmayer}},
  \bibinfo {author} {\bibfnamefont {K.}~\bibnamefont {Starke}}, \bibinfo
  {author} {\bibfnamefont {S.}~\bibnamefont {Gorovikov}}, \bibinfo {author}
  {\bibfnamefont {J.~E.}\ \bibnamefont {Prieto}}, \bibinfo {author}
  {\bibfnamefont {K.}~\bibnamefont {D\"obrich}}, \bibinfo {author}
  {\bibfnamefont {S.}~\bibnamefont {Bl\"ugel}},\ and\ \bibinfo {author}
  {\bibfnamefont {G.}~\bibnamefont {Kaindl}},\ }\href
  {https://doi.org/10.1103/PhysRevB.71.201403} {\bibfield  {journal} {\bibinfo
  {journal} {Phys. Rev. B}\ }\textbf {\bibinfo {volume} {71}},\ \bibinfo
  {pages} {201403} (\bibinfo {year} {2005})}\BibitemShut {NoStop}%
\bibitem [{\citenamefont {LaShell}\ \emph {et~al.}(1996)\citenamefont
  {LaShell}, \citenamefont {McDougall},\ and\ \citenamefont
  {Jensen}}]{bib:lashell96}%
  \BibitemOpen
  \bibfield  {author} {\bibinfo {author} {\bibfnamefont {S.}~\bibnamefont
  {LaShell}}, \bibinfo {author} {\bibfnamefont {B.~A.}\ \bibnamefont
  {McDougall}},\ and\ \bibinfo {author} {\bibfnamefont {E.}~\bibnamefont
  {Jensen}},\ }\href {https://doi.org/10.1103/PhysRevLett.77.3419} {\bibfield
  {journal} {\bibinfo  {journal} {Phys. Rev. Lett.}\ }\textbf {\bibinfo
  {volume} {77}},\ \bibinfo {pages} {3419} (\bibinfo {year}
  {1996})}\BibitemShut {NoStop}%
\bibitem [{\citenamefont {Petersen}\ and\ \citenamefont
  {Hedegård}(2000)}]{bib:petersen00}%
  \BibitemOpen
  \bibfield  {author} {\bibinfo {author} {\bibfnamefont {L.}~\bibnamefont
  {Petersen}}\ and\ \bibinfo {author} {\bibfnamefont {P.}~\bibnamefont
  {Hedegård}},\ }\href
  {https://doi.org/https://doi.org/10.1016/S0039-6028(00)00441-6} {\bibfield
  {journal} {\bibinfo  {journal} {Surface Science}\ }\textbf {\bibinfo {volume}
  {459}},\ \bibinfo {pages} {49} (\bibinfo {year} {2000})}\BibitemShut
  {NoStop}%
\bibitem [{\citenamefont {Abate}\ and\ \citenamefont
  {Asdente}(1965)}]{bib:abate65}%
  \BibitemOpen
  \bibfield  {author} {\bibinfo {author} {\bibfnamefont {E.}~\bibnamefont
  {Abate}}\ and\ \bibinfo {author} {\bibfnamefont {M.}~\bibnamefont
  {Asdente}},\ }\href {https://doi.org/10.1103/PhysRev.140.A1303} {\bibfield
  {journal} {\bibinfo  {journal} {Phys. Rev.}\ }\textbf {\bibinfo {volume}
  {140}},\ \bibinfo {pages} {A1303} (\bibinfo {year} {1965})}\BibitemShut
  {NoStop}%
\end{thebibliography}%


\begin{thebibliography}{0}%
\makeatletter
\providecommand \@ifxundefined [1]{%
 \@ifx{#1\undefined}
}%
\providecommand \@ifnum [1]{%
 \ifnum #1\expandafter \@firstoftwo
 \else \expandafter \@secondoftwo
 \fi
}%
\providecommand \@ifx [1]{%
 \ifx #1\expandafter \@firstoftwo
 \else \expandafter \@secondoftwo
 \fi
}%
\providecommand \natexlab [1]{#1}%
\providecommand \enquote  [1]{``#1''}%
\providecommand \bibnamefont  [1]{#1}%
\providecommand \bibfnamefont [1]{#1}%
\providecommand \citenamefont [1]{#1}%
\providecommand \href@noop [0]{\@secondoftwo}%
\providecommand \href [0]{\begingroup \@sanitize@url \@href}%
\providecommand \@href[1]{\@@startlink{#1}\@@href}%
\providecommand \@@href[1]{\endgroup#1\@@endlink}%
\providecommand \@sanitize@url [0]{\catcode `\\12\catcode `\$12\catcode
  `\&12\catcode `\#12\catcode `\^12\catcode `\_12\catcode `\%12\relax}%
\providecommand \@@startlink[1]{}%
\providecommand \@@endlink[0]{}%
\providecommand \url  [0]{\begingroup\@sanitize@url \@url }%
\providecommand \@url [1]{\endgroup\@href {#1}{\urlprefix }}%
\providecommand \urlprefix  [0]{URL }%
\providecommand \Eprint [0]{\href }%
\providecommand \doibase [0]{https://doi.org/}%
\providecommand \selectlanguage [0]{\@gobble}%
\providecommand \bibinfo  [0]{\@secondoftwo}%
\providecommand \bibfield  [0]{\@secondoftwo}%
\providecommand \translation [1]{[#1]}%
\providecommand \BibitemOpen [0]{}%
\providecommand \bibitemStop [0]{}%
\providecommand \bibitemNoStop [0]{.\EOS\space}%
\providecommand \EOS [0]{\spacefactor3000\relax}%
\providecommand \BibitemShut  [1]{\csname bibitem#1\endcsname}%
\let\auto@bib@innerbib\@empty
\end{thebibliography}%

\end{document}



\title{Supplemental Material for: 
``Nature of Interfacial Dzyaloshinskii-Moriya Interactions in Graphene/Co/Pt(111) Multilayer Heterostructures''}



\newcommand{\QUIMI}[0]{{
Departamento de Pol\'{\i}meros y Materiales Avanzados: 
F\'{\i}sica, Qu\'{\i}mica y Tecnolog\'{\i}a, Facultad de Qu\'{\i}mica UPV/EHU,
Apartado 1072, 20080 Donostia-San Sebasti\'an, Spain}}

\newcommand{\DIPC}[0]{{
Donostia International Physics Center, 
Paseo Manuel de Lardiz\'abal 4, 20018 Donostia-San Sebasti\'an, Spain}}

\newcommand{\CFM}[0]{{
Centro de F\'{\i}sica de Materiales CFM/MPC (CSIC-UPV/EHU), 
Paseo Manuel de Lardiz\'abal 5, 20018 Donostia-San Sebasti\'an, Spain}}

\newcommand{\JUELICH}[0]{{
Peter Gr\"unberg Institut and Institute for Advanced Simulation, 
Forschungszentrum J\"ulich and JARA, 52425 J\"ulich, Germany}}

\newcommand{\ICMM}[0]{{
Instituto de Ciencia de Materiales de Madrid, CSIC,
Cantoblanco, 28049 Madrid, Spain}}


\author{M. Blanco-Rey}
\affiliation{\QUIMI}
\affiliation{\DIPC}
\author{G. Bihlmayer}
\affiliation{\JUELICH}
\author{A. Arnau}
\affiliation{\CFM}
\affiliation{\QUIMI}
\affiliation{\DIPC}
\author{J.I. Cerd\'a}
\affiliation{\ICMM}


\date{\today}



\maketitle

\newpage

\begin{figure}[tb!]
\centerline{\includegraphics[width=0.5\columnwidth]{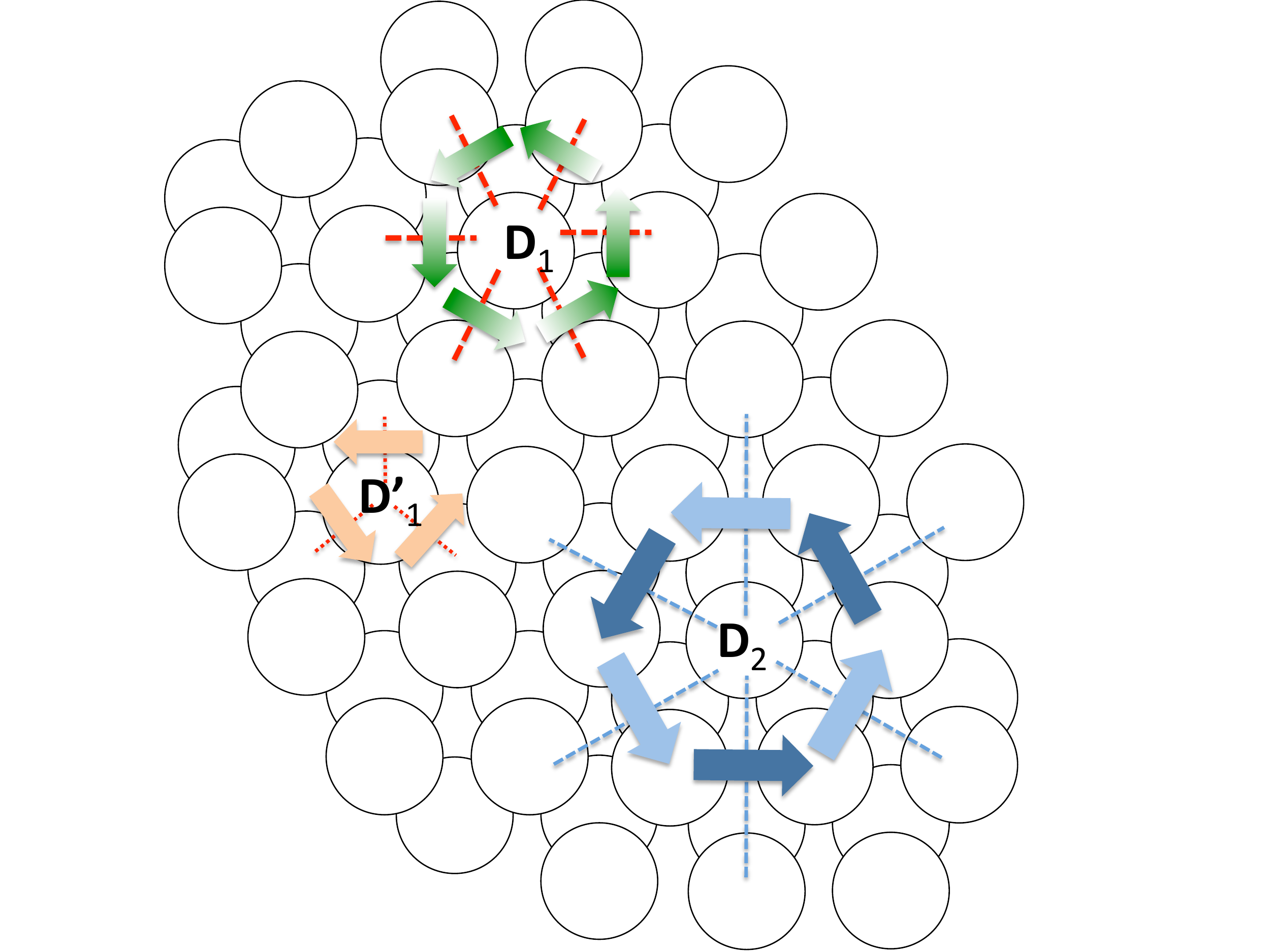}
\includegraphics[width=0.5\columnwidth]{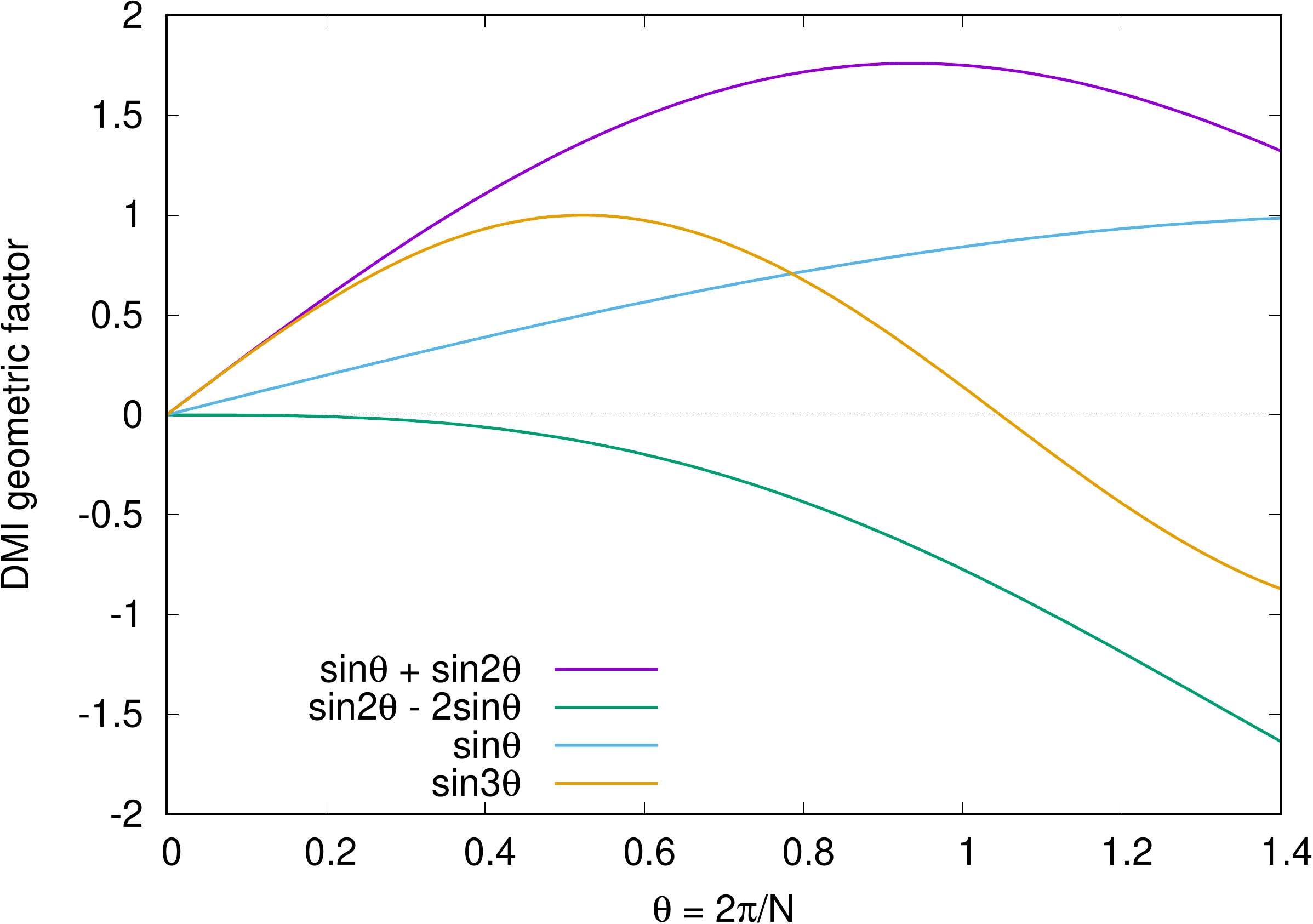}} 
\caption{\textbf{Short-ranged DMI contributions on a 3-fold 
symmetric bilayer.} 
Left: $D$-vectors at a $fcc$ bilayer for first ($\mathbf{D}_1$) and second ($\mathbf{D}_2$) 
in-plane nearest neighbours, and
first nearest neighbours of consecutive atomic planes ($\mathbf{D}'_1$).
The colour gradient from green to white in $\mathbf{D}'_1$ depicts the variation 
of the $D_{1z}$ component sign. $\mathbf{D}_2$ and $\mathbf{D}'_1$ do not have a 
perpendicular component. $\mathbf{D}_2$ is 3-fold symmetric with no perpendicular component, 
i.e. there are two different vectors for the second nearest neighbour interaction.
Right: for spin spirals defined by $q$-vectors $\mathbf{q} = \frac{2\pi}{a} ( \frac{1}{N}, \frac{1}{N}, 0)$,
the in plane first nearest neighbours contribution is 
$\Delta E_{DMI,1}^Y (\theta) = 4 S^2 D_{1y} ( \sin \theta + \sin 2\theta ) $ 
and $\Delta E_{DMI,1}^Z (\theta) = 4 S^2 D_{1z} ( \sin 2\theta - 2\sin \theta ) $,
where $S$ is the atomic spin, $\theta = \frac{2\pi}{N}$.
Each surface atom has three nearest neighbours at the plane below,
which contribute as $\Delta E_{DMI,1'}^Y \theta) = 2 SS' D'_{1x} \sqrt{3} \sin \theta  $,
where $S'$ is the atomic spin in the second plane.
For the second in-plane nearest neighbours, the energy term is 
$\Delta E_{DMI,2}^Y \theta) = 2 S^1 (D_{2x}-\tilde D_{2x}) \sqrt{3} \sin 3\theta  $, 
noting that the interaction is three-fold symmetric.
The figure shows the angle-dependence of the geometric factors in these energy terms.
}
\label{fig:geom_Ds}
\end{figure}


\begin{figure}[tb!]
\centerline{\includegraphics[width=\columnwidth]{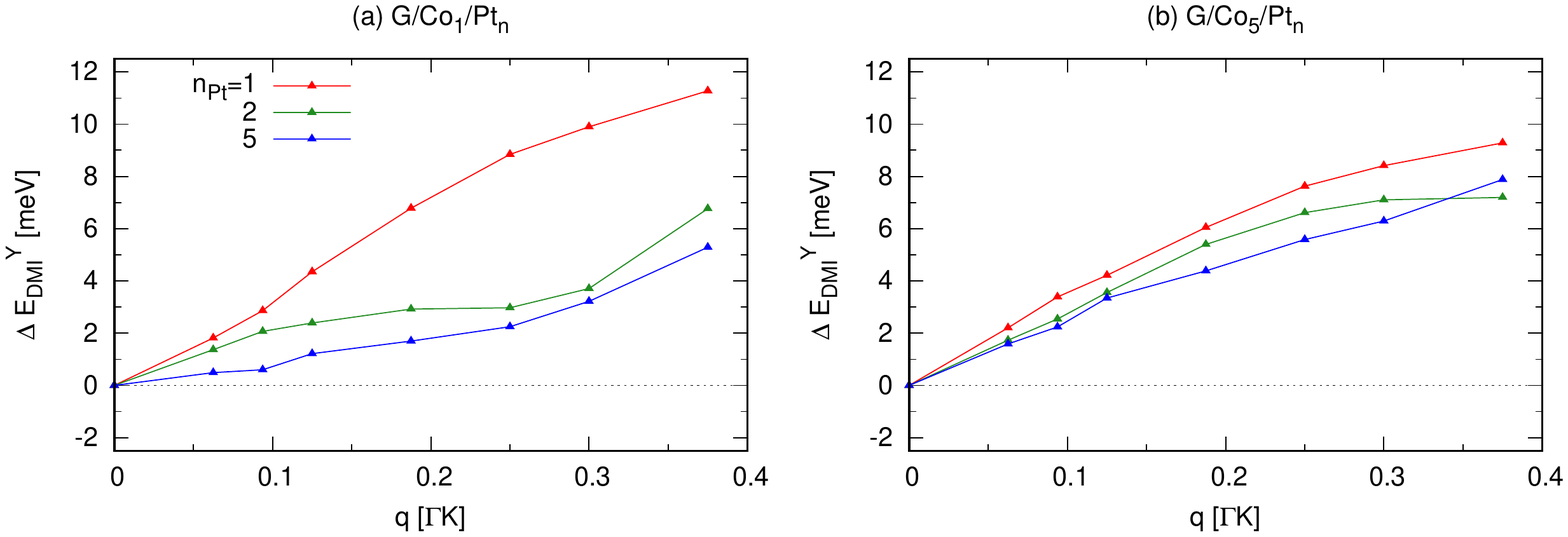}}
\caption{\textbf{DMI energy dependence on the Pt substrate thickness.}
DMI energy $\Delta E_{DMI}^{Y}(\mathbf{q})$ dependence on the Pt
substrate thickness for one (a) and five (b) Co layers.
Here, too, we observe the aforementioned substrate effect in the Co monolayer limit:
there is a noticeable change in the Gr/Co$_1$/Pt$_n$ behaviour when
a second Pt plane is added, that brings the energy close to the thick Pt
substrate limit (a). This crossover is not visible in the thick Co layer limit.
}
\label{fig:pt_thickness}
\end{figure}


\begin{figure}[tb!]
\centerline{\includegraphics[width=\columnwidth]{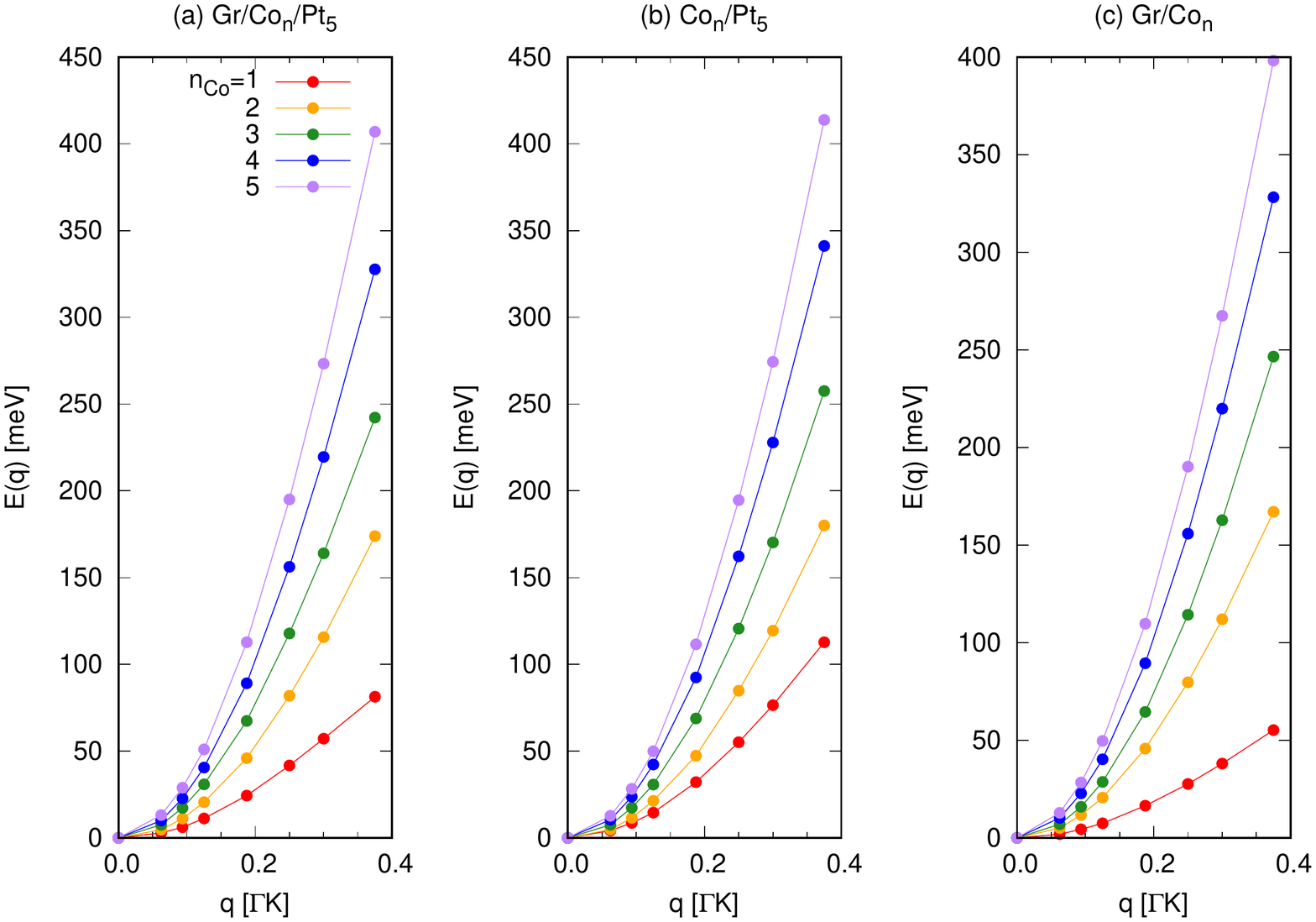}}
\caption{\textbf{Energy dispersion of the spin spirals in Gr/Co$_n$/Pt$_5$.}
Dependence on the Co layer thickness of the spin spiral energy dispersion $E(\mathbf{q})$
in Gr/Co$_n$Pt$_5$ slabs and their constituents  Co$_n$/Pt$_5$ and Gr/Co$_n$ .
These values are calculated for the $q$-vector along the $\Gamma K$ direction without SOC
and in a non-self-consistent scheme.
}
\label{fig:Eq}
\end{figure}


\begin{figure}[tb!]
\centerline{\includegraphics[width=\columnwidth]{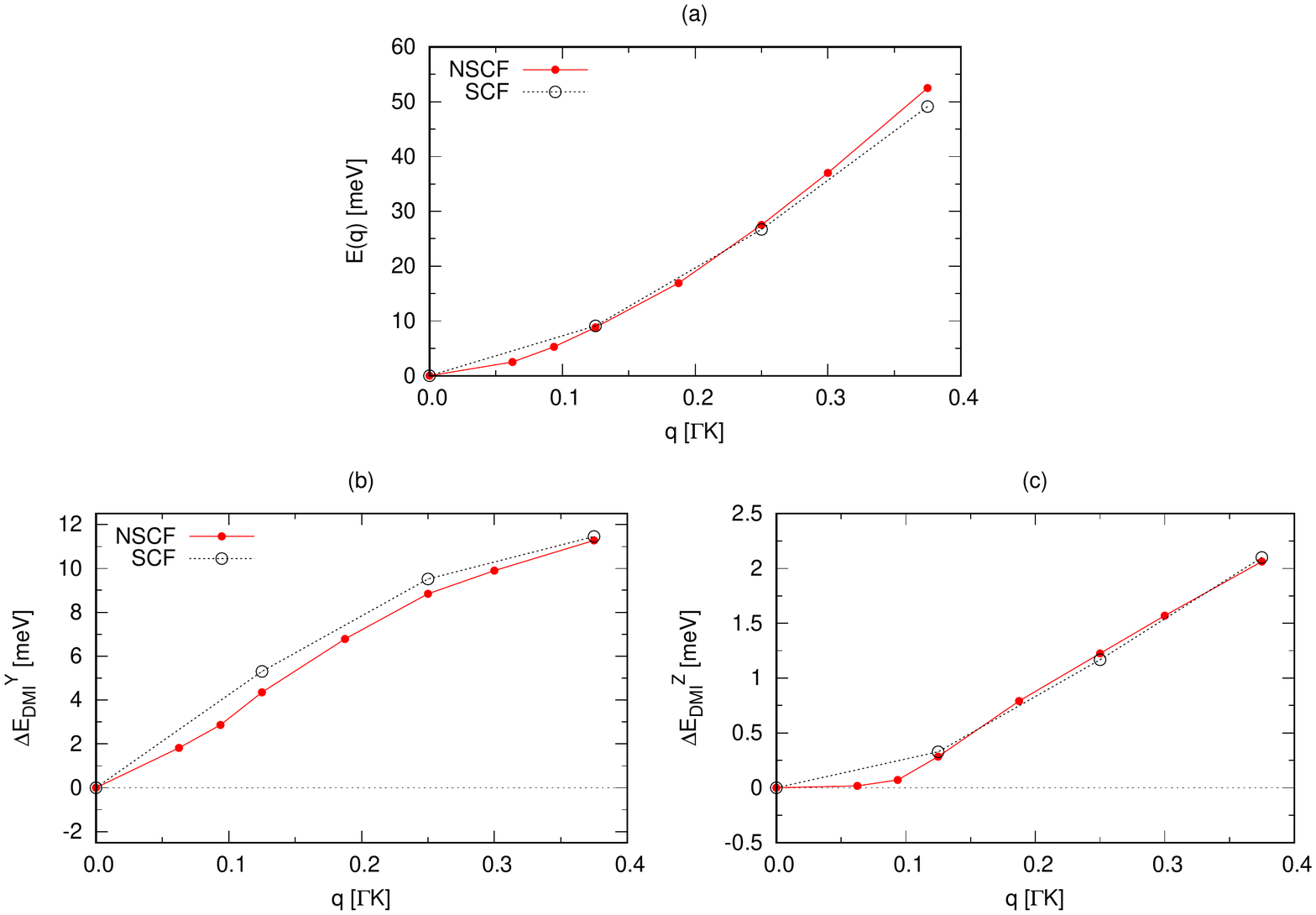}}
\caption{\textbf{Effect of self-consistency in the spin spiral calculation of the Gr/Co$_1$/Pt$_1$ slab.}
(a) Spin spiral energy dispersion in the absence of SOC for a Gr/Co$_1$/Pt$_1$
trilayer, with the $q$-vector along the $\Gamma K$ direction. 
The red filled dots are calculated by imposing the non-collinear magnetization density 
given by the Generalized Bloch Theorem (GBT) for each $q$-vector on the 
collinear electron wavefunctions, i.e. the spin spiral energy is 
determined non-self-consistently (NSCF) or as a perturbation. 
The energies of optimized spirals after reaching self-consistency (SCF)
in the Kohn-Sham equations are given by the black empty dots.
At the low-$q$ regime used in this work, agreement is good.
(b,c) Comparison of the 
DMI energies $\Delta E_{DMI}^{\hat{s}_a}(\mathbf{q})$ for $\hat{s}_a=Y$ (b) and $Z$ (c)
calculated by adding SOC perturbatively to the NSCF and SCF spirals 
obtained above. We observe that the agreement is quantitative, and  
thus the use of a perturbative approach to obtain the spinors from the 
collinear electron density of the system without SOC is justified in this $q$ value range. 
Note that achieving self-consistency would be too computationally demanding
for the more realistic modes with five Pt planes as a substrate.
}
\label{fig:nscf_vs_scf}
\end{figure}


\begin{figure}[tb!]
\centerline{\includegraphics[width=\columnwidth]{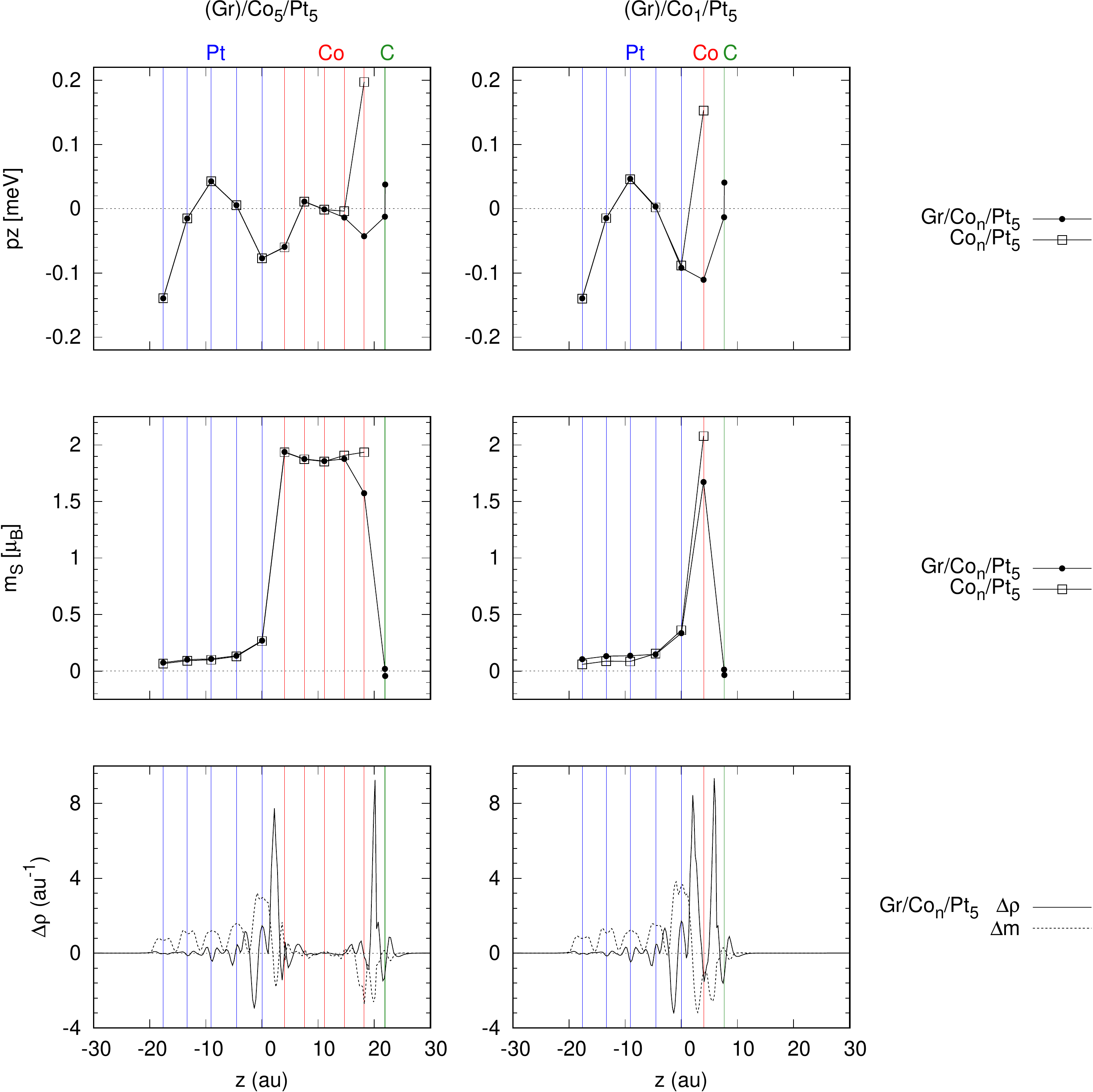}}
\caption{\textbf{Dipoles, spin polarization and charge distribution 
at the interfaces.}
From top to bottom, perpendicular dipole $p_z$ in the muffin tin, 
spin moment in the muffin tin $m_S$ and
charge density difference $\Delta \rho (z)$ for the 
Gr/Co$_5$/Pt$_5$ (left column) and Gr/Co$_5$/Pt$_5$ (right) systems.
Dots and squares show the data with and without graphene capping, 
respectively. 
In the bottom panels, the $\Delta \rho(z)$ quantity (solid line) is 
calculated as the $xy$-integrated substraction of charge densities
$\rho (\mathrm{Gr}/\mathrm{Co}_n/\mathrm{Pt}_5) - 
 \rho (\mathrm{Gr}) - \rho (\mathrm{Co}_n) - \rho (\mathrm{Pt}_5) $,
which accounts for charge redistribution at both Gr/Co and Co/Pt
interfaces. 
The dashed line $\Delta m(z)$  shows the magnetization density for 
$\Delta \rho(z)$.
}
\label{fig:cdd}
\end{figure}


\begin{figure}[tb!]
\centerline{\includegraphics[width=\columnwidth]{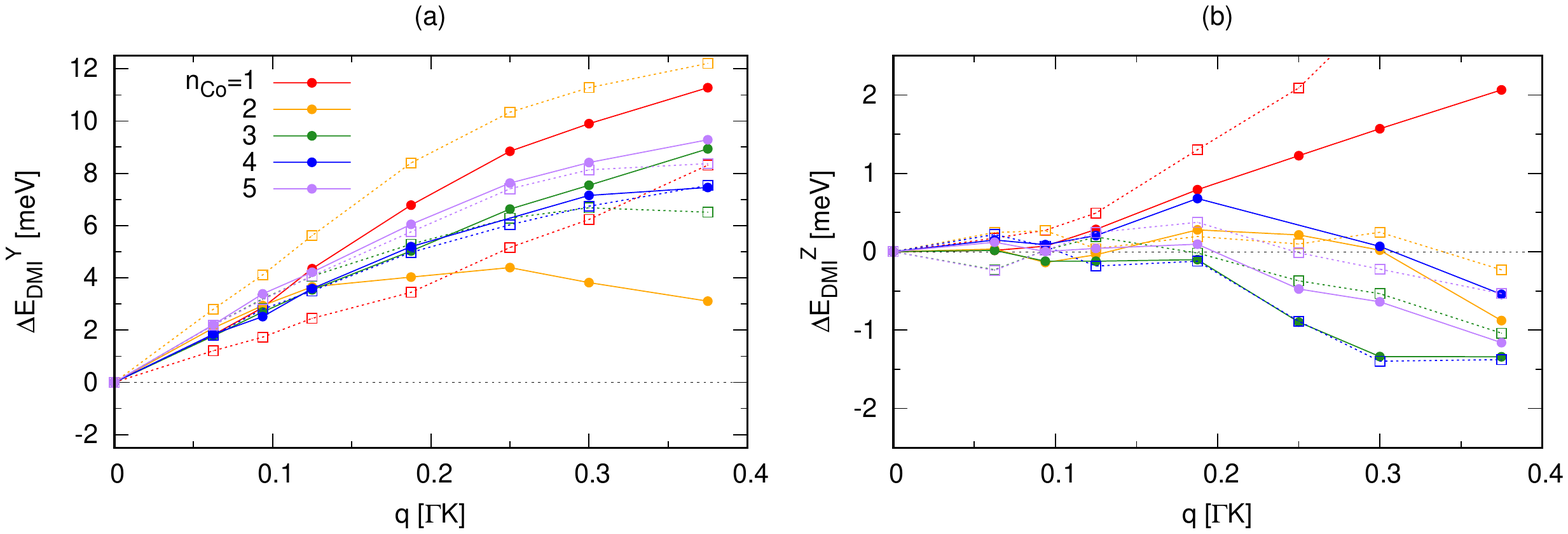}}
\caption{\textbf{DMI energy for Gr/Co$_n$/Pt$_1$ slabs constituents.}
The filled dots show the
dependence of the DMI energy $\Delta E_{DMI}^{\hat{s}_a}(\mathbf{q})$
on the Co thickness for model slabs with a single Pt monolayer.
Each color corresponds to a thickness.
The sum of the energies of the Gr/Co$_n$ and Co$_n$/Pt$_1$  slab constituents
is indicated by empty squares to check interfacial DMI additivity at each thickness.
On panel (a) for $\hat{s}_a=Y$ three observations are made:
(i) additivity is not met for $n=1,2$ yet;
(ii) DMI energy values converge from $n=3$ after oscillating; and
(iii) the $n=2$ curve has a maximum at
$q \approx 0.2 |\Gamma K|$ due to the effect of
the $\Delta E_{DMI,1'}^Y$ contribution (see Fig.~\ref{fig:geom_Ds}),
which inverts the sign of the $\sin\theta$ leading term in the DMI energy.
The same results are reported in the main paper with a Pt$_5$ substrate.
However, energies for $n=1,2$ differ (they are about a factor two larger in
Gr/Co$_1$/Pt$_1$ than in Gr/Co$_1$/Pt$_5$ and a factor three smaller
in Gr/Co$_2$/Pt$_1$ than in Gr/Co$_2$/Pt$_5$), proving that the
ultrathin Co regime is strongly dependent on the electronic structure
of the Pt substrate and the DMI is not localized at the precise Co-Pt interface.
Instead, it depends on the hybridization between Co and deeper Pt planes.
In panel (b) for $\hat{s}_a=Z$ we observe an energy change from $n=1$ to thicker
Co layers, i.e. the sign of $D_z$ is inverted.
Note that this sign change is not obtained on the Pt$_5$ substrate.
}
\label{fig:partialslabs_pt1}
\end{figure}


\begin{figure}[tb!]
\centerline{\includegraphics[width=\columnwidth]{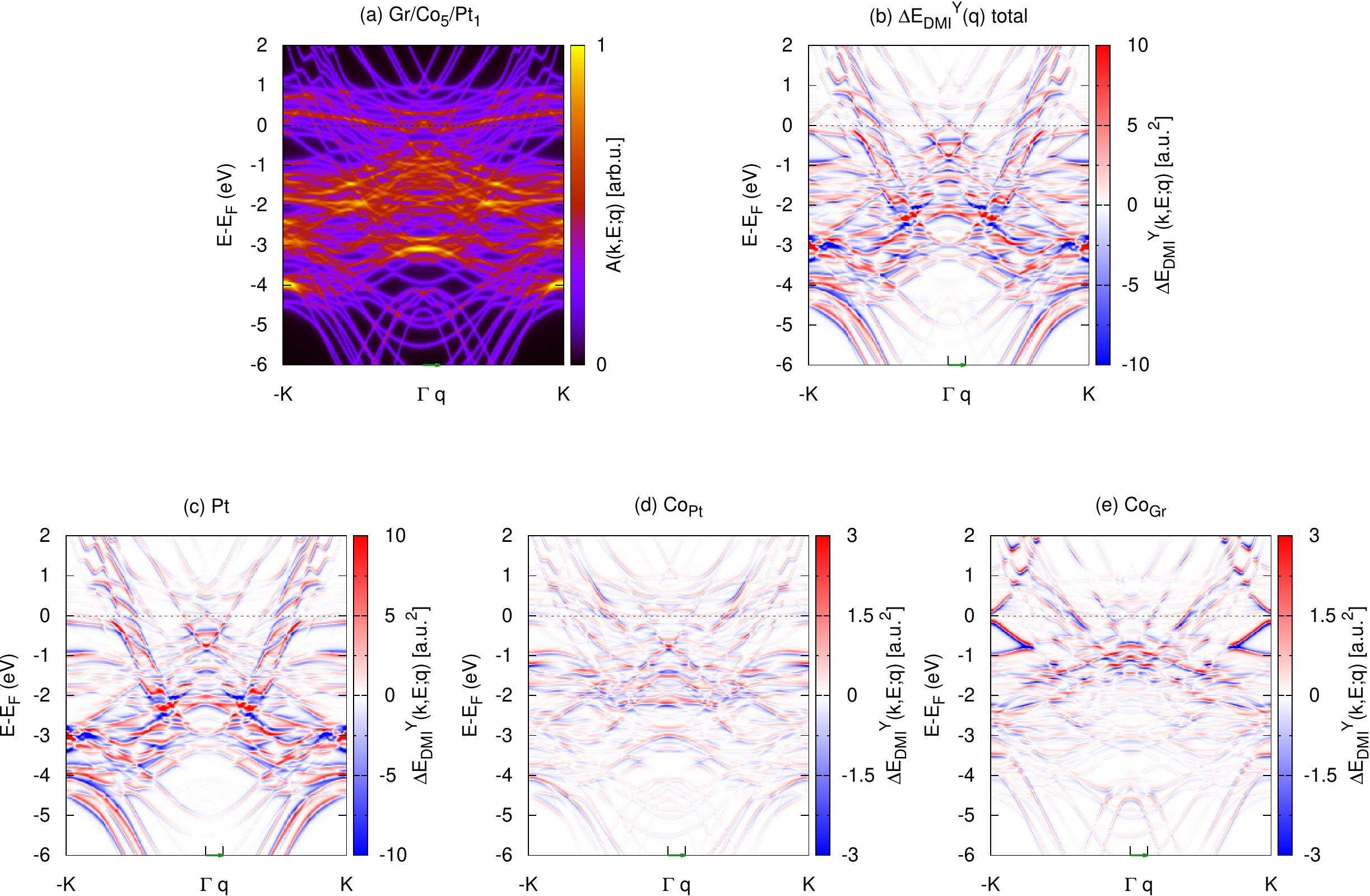}}
\caption{\textbf{$k$-resolved spectral DMI energy for Gr/Co$_5$/Pt$_1$.}
(a) Spectral density of the Gr/Co$_5$/Pt$_1$ system calculated without SOC for the 
$q$-vector $\mathbf{q} = \frac{2\pi}{a} ( \frac{1}{24}, \frac{1}{24}, 0)$ 
(q=$0.125|\Gamma K|$), indicated with a green arrow. 
The lateral splitting introduced by the spiral in the bands is observed.
(b) Density of energy $\Delta E_{DMI}^{Y}(\mathbf{q})$.
The actual DMI energy is the integral over occupied states, 
which involves many cancellations of positive and negative contributions.
(c-e) The contributions of the indicated interfacial atoms to the density 
of panel (b). The more intense DMI contributions in panels (c,d) map 
the broad bands mainly originated from Pt($d$) and Co($d$) orbitals,
centred at $E_F-3.5$ and $E_F-1.5$\,eV, respectively. 
The Co$_\mathrm{Gr}$ interfacial states show high intensity at the 
Co-C hybrid conical bands. Nevertheless, as in the other panels, 
cancellations of DMI energies of opposite sign attenuate 
the contribution of this band structure feature to the total DMI energy.
}
\label{fig:bands_ptco1g}
\end{figure}


\pagebreak